\documentclass{article}
\usepackage{bm}
\usepackage{amsmath,amssymb}
\usepackage[top=20truemm,bottom=20truemm,left=20truemm,right=20truemm]{geometry}
\usepackage{authblk}
\usepackage[subrefformat=parens]{subcaption} 
\usepackage{siunitx} 
\usepackage{soul} 
\usepackage{mdframed}
\usepackage[eprint=true,backend=biber,style=phys,articletitle=false,biblabel=brackets,chaptertitle=false,pageranges=false,uniquelist=false,maxbibnames=2,arxiv=abs]{biblatex}
\usepackage{hyperref}
\usepackage{xurl}
\usepackage{breqn} 
\hypersetup{breaklinks=true}
\DeclareSourcemap{
	\maps[datatype=bibtex, overwrite=true]{
		\map{
			\step[fieldset=doi, null]
			\step[fieldsource=journal,
				match={Physical Review D},
				replace={Phys. Rev. D}
			]
			\step[fieldsource=journal,
				match={Philosophical Transactions of the Royal Society A},
				replace={Philos. T. R. Soc. A}
			]
			\step[fieldsource=journal,
				match={The Astrophysical Journal Letters},
				replace={Astrophys. J. Lett.}
			]
			\step[fieldsource=journal,
				match={The Astrophysical Journal},
				replace={Astrophys. J.}
			]
			\step[fieldsource=journal,
				match={Classical and Quantum Gravity},
				replace={Classical Quant. Grav.}
			]
			\step[fieldsource=journal,
				match={Journal of High Energy Physics},
				replace={JHEP}
			]
			\step[fieldsource=journal,
				match={Proceedings of the American Mathematical Society},
				replace={P. Am. Math. Soc.}
			]
			\step[fieldsource=journal,
				match={Journal of Physics A: Mathematical and Theoretical},
				replace={J. Phys. A-Math. Theor.}
			]
			\step[fieldsource=journal,
				match={Journal of Physics A: Mathematical and General},
				replace={J. Phys. A-Math. Gen.}
			]
			\step[fieldsource=journal,
				match={Inventiones mathematicae},
				replace={Invent. Math.}
			]
			\step[fieldsource=journal,
				match={Journal of Geometry and Physics},
				replace={J. Geom. Phys.}
			]
			\step[fieldsource=journal,
				match={Journal of the Faculty of Science, the University of Tokyo. Sect. 1 A, Mathematics},
				replace={J. Fac. Sci. U. Tokyo 1}
			]
			\step[fieldsource=journal,
				match={Compositio Mathematica},
				replace={Compos. Math.}
			]
			\step[fieldsource=journal,
				match={Proceedings of the Japan Academy, Series A, Mathematical Sciences},
				replace={P. Jpn. Acad. A-Math.}
			]
			\step[fieldsource=journal,
				match={Journal of Computational and Applied Mathematics},
				replace={J. Comput. Appl. Math.}
			]
			\step[fieldsource=journal,
				match={Constructive Approximation},
				replace={Constr. Approx.}
			]
			\step[fieldsource=journal,
				match={The European Physical Journal C},
				replace={Eur. Phys. J. C}
			]
			\step[fieldsource=journal,
				match={General Relativity and Gravitation},
				replace={Gen. Relativ. Gravit.}
			]
			\step[fieldsource=journal,
				match={American Journal of Physics},
				replace={Am. J. Phys.}
			]
		}
	}
}

\makeatletter
	
		\@addtoreset{equation}{section}
\makeatother
\title{Strong gravitational lensing by a Reissner-Nordstr\"{o}m naked singularity with a marginally unstable photon sphere}
\date{\today}
\author{Tadashi Sasaki\thanks{ta-sasaki@kumagaku.ac.jp}}
\affil{Faculty of Commerce, Kumamoto Gakuen University, Kumamoto 862-8680, Japan}
\begin{document}

\maketitle

\begin{abstract}
{{{
	We investigate strong gravitational lensing by a marginally unstable photon sphere in a Reissner-Nordstr\"{o}m naked singularity spacetime.
	Using the Picard-Fuchs equation, we derive full-order power series expressions for the deflection angle in various regimes, including
	the strong deflection limits from both outside and inside the photon sphere.
	We show that the deflection angle diverges non-logarithmically in both cases, refining existing asymptotic formulae.
	Comparing truncated approximations with numerical results, we find that higher-order corrections are essential to achieve
	comparable accuracy to logarithmic divergence cases.
	Using these improved formulae, we also derive precise approximations for image positions that are not restricted to the almost 
	perfectly aligned cases.
}}}
\end{abstract}


\section{Introduction}
{{{
	Over 100 years after the confirmation of general relativity by a total solar eclipse in 1919\cite{Dyson_1920},
	the bending of light by massive bodies has been an indispensable phenomenon to observe our universe.
	In particular, the Event Horizon Telescope (EHT) collaboration has succeeded to image the shadow due to the black hole candidates
	at the centers of galaxies\cite{Akiyama_2019,Akiyama_2022}.

	It has been known that spacetime models for static and spherically symmetric massive compact objects possess a family of circular 
	null geodesics, which is called a photon sphere or an anti-photon sphere depending on the stability of the geodesics.
	If a photon incident on such a massive body passes close to the photon sphere, it goes around the body many times before being scattered to infinity.
	As a consequence, a single source of light can yield an infinite number of images, which are called relativistic images\cite{VirbhadraEllis_2000}.

	One of the fundamental quantities for gravitational lensing is the deflection angle $\alpha$ 
	for a photon incident from and scattered to infinity.
	In a static and spherically symmetric spacetime, null geodesics incident from infinity can be characterized by the impact parameter $b$,
	which is defined by the distance between the line of sight and the asymptote of the geodesic.
	The existence of the relativistic images for a spacetime with a photon sphere is reflected in the divergence of the deflection angle 
	at some critical value of the impact parameter.
	Bozza\cite{Bozza_2002} showed that this divergence can typically be approximated by
	\begin{equation}
		\alpha\sim-\bar{a}\log\left(\frac{b}{b_{\rm c}}-1\right)+\bar{b},\ b\to b_{\rm c} \label{SDLBozza}
	\end{equation}
	where $\bar{a}$ and $\bar{b}$ are constants determined by the spacetime metric and $b_{\rm c}(>0)$ is the critical impact parameter.
	When a photon comes with the impact parameter $b=b_{\rm c}$, it will be captured by the photon sphere.
	The limit $b\to b_{\rm c}$ is called the strong deflection limit (SDL).
	In order to distinguish various models for massive compact objects, 
	considerable effort has been made to calculate the coefficients $\bar{a}$ and $\bar{b}$ for black holes\cite{Eiroa_2002,Gyulchev2006,Gyulchev2007,GhoshSenGupta2010,ChibaKimura2017,TsukamotoGong_2017,Tsukamoto_2017,Hsieh_etal2021,Ghosh_2022,Soares2023,Junior_2024,Soares2024},
	naked singularities\cite{Paul2020,Tsukamoto_2021}, 
	wormholes\cite{Tsukamoto2021,Soares2023b,Hsieh_2025}, and so on.
	We also note that it has recently been shown that the coefficient $\bar{a}$ can be written in terms of the curvature of the spacetime\cite{Igata2025}.

	Recently, however, several singular cases where this logarithmic formula fails to approximate $\alpha$ in the SDL\cite{Paul2020,Tsukamoto2020b,Tsukamoto2020,Tsukamoto2021}.
	In each case, the spacetime has a marginally unstable photon sphere, which appears due to the confluence of photon and anti-photon spheres,
	and the asymptotic behavior of the deflection angle in the SDL is given by
	\begin{equation}
		\alpha\sim\frac{\bar{c}}{(b/b_{\rm m}-1)^\lambda}+\bar{d},\ b\to b_{\rm m} \label{SDLpower}
	\end{equation}
	where $\lambda$, $\bar{c}$ and $\bar{d}$ are constants and 
	the critical impact parameter $b_{\rm m}$ corresponds to the orbit trapped by the marginally unstable photon sphere.
	We also note that a similar behavior happens when a photon sphere and the event horizon merge as shown for the Kerr spacetime in \cite{Barlow_2017}.

	In this paper, we consider gravitational lensing in the Reissner-Nordstr\"{om} (RN) spacetime,
	which possesses mass $M(>0)$ and electric charge $Q\in\mathbb{R}$.
	The strong deflection limit in the RN black hole spacetime was first treated in \cite{Eiroa_2002}, where the constants $\bar{a}$ and $\bar{b}$ were calculated numerically for various values of electric charge.
	After that Bozza\cite{Bozza_2002} explicitly derived the charge dependence of $\bar{a}$ although only the 2nd-order terms with respect to the electric charge were given for $\bar{b}$.
	The exact expression for $\bar{b}$ was obtained in\cite{TsukamotoGong_2017,Tsukamoto_2017} and was recovered in \cite{Sasaki_2024}.
	On the other hand, the weak deflection limit $b\to+\infty$ has been intermittently studied in, for example, \cite{Eiroa_2002,Keeton_2005,Jia2020},
	and recently a full-order expression has been obtained in \cite{Sasaki_2024}.

	If the electric charge of the RN spacetime exceeds a critical value, namely for $Q^2>M^2$\footnote{We employ the geometrized units $c=G=1$.}, 
	it is known that the spacetime loses the event horizon to become a naked singularity.
	Even in that case, as long as the charge is less than another critical value, namely $1<Q^2/M^2\leq9/8$,
	the spacetime has a photon sphere and anti-photon sphere\footnote{When $Q^2/M^2\leq1$, the anti-photon sphere is hidden by the event horizon.}.
	Tsukamoto\cite{Tsukamoto2020,Tsukamoto_2021} investigated strong gravitational lensing by a RN naked singularity.
	In this case, in addition to the usual SDL, which is approached from outside the photon sphere $b\to b_{\rm c}+0$, 
	one can consider the SDL from inside the photon sphere, i.e. $b\to b_{\rm c}-0$.
	For non-singular cases $1<Q^2/M^2<9/8$, the standard formula (\ref{SDLBozza}) is applicable for the SDL from outside.
	In the SDL from inside the photon sphere, 
	Shaikh et al.\cite{Shaikh2019} showed that, in a general static and spherically symmetric spacetime, the deflection angle again diverges logarithmically,
	\begin{equation}
		\alpha\sim-\bar{a}\log\left(\frac{b_c^2}{b^2}-1\right)+\bar{b},\ b\to b_c-0,
	\end{equation}
	where the coefficients $\bar{a},\bar{b}$ are different from those in (\ref{SDLBozza}).
	In this paper, ovservables in the RN naked singularity for a specific value $Q^2/M^2=1.05$ was calculated as an example.
	Tsukamoto\cite{Tsukamoto_2021} then derived explicit expressions for $\bar{a},\bar{b}$ as functions of $Q/M$
	for both the outside and inside cases.
	At the same time, these logarithmic formulae were shown to break down in the marginal limit $Q^2/M^2\to9/8-0$,
	where the photon sphere and anti-photon sphere merge to form a marginally unstable photon sphere.
	The singular case $Q^2/M^2=9/8$ was partially treated in \cite{Tsukamoto2020} and he argued that 
	for the SDL from outside the photon sphere, the deflection angle can be approximated by (\ref{SDLpower}) with
	\begin{equation}
		\lambda=\frac{1}{6},\ \  \bar{c}=2^{5/3}\cdot3^{1/2},\ \  \bar{d}=-\sqrt{6}-\pi.\label{cf_Tsukamoto2020}
	\end{equation}
	However, the corresponding formula for the SDL from inside the photon sphere is missing.

	To derive the missing formula, we take a different approach, which has been applied to the cases of the Schwarzschild and 
	RN black holes in \cite{Sasaki_2021,Sasaki_2024}.
	In the literature, the SDL analysis of the deflection angle has usually been performed by directly expanding integral representations of $\alpha$.
	On the other hand, we solve the Picard-Fuchs (PF) equation satisfied by the deflection angle.
	In the RN spacetime, the deflection angle can be expressed as an elliptic integral, and
	the PF equation is then given by a 2nd-order linear differential equation.
	Although only the black hole case, i.e. $0<Q^2/M^2\leq1$ was focused in \cite{Sasaki_2024}, 
	the derived PF equation is in fact valid for the naked singularity case $1<Q^2/M^2$.
	Compared with directly expanding the integral representation, solving the linear differential equation is advantageous
	to derive higher order corrections since one can obtain recurrence relations for the expansion coefficients from the differential equations.

	This paper is organized as follows: in section \ref{sec_intrep}, we derive an integral representation of the deflection angle 
	and show that it can be seen as a function of $x=27M^2/2b^2$. 
	In section \ref{sec_PFeq}, we derive four exact expressions in terms of various hypergeometric series to fully analyze the deflection angle
	in the whole region of $x>0$.
	In particular, it is shown that $\alpha$ in the SDL from the inside has an asymptotic form similar to (\ref{SDLpower}).
	Section \ref{sec_accuracy} is devoted to investigating the accuracy of the derived power series expansions.
	We argue that the asymptotic formula (\ref{SDLpower}) and the similar one for the SDL from the inside are 
	less accurate than the standard logarithmic formula (\ref{SDLBozza})
	and that it is necessary to include higher-order terms.
	We also compare our result for the SDL from outside the photon sphere with that of \cite{Tsukamoto2020}.
	In section \ref{sec_lenseq}, we solve the lens equation using the approximation for the deflection angle derived in the previous sections.
	In particular, we calculate the diameters of the relativistic Einstein rings both outside and inside the photon sphere.
	We summarize the paper and give some discussions in section \ref{sec_discussion}.

}}}

\section{Integral Representation of the Deflection Angle\label{sec_intrep}}
{{{
	We consider the RN metric,
	\begin{equation}
		\mathrm{d}s^2=-\left(1-\frac{2M}{r}+\frac{Q^2}{r^2}\right)\mathrm{d}t^2+\left(1-\frac{2M}{r}+\frac{Q^2}{r^2}\right)^{-1}\mathrm{d}r^2+r^2\left(\mathrm{d}\theta^2+\sin^2\theta\mathrm{d}\phi^2\right),
	\end{equation}
	where $M (>0)$ and $Q\in\mathbb{R}$ represent the mass and the electric charge of the spacetime respectively.
	In this paper, we restrict our consideration to the case of $Q^2=9M^2/8$, where the spacetime represents an isolated naked singularity
	with a marginally unstable photon sphere at the radius $r=r_{\rm ps}=3M/2$.

	Light ray travels along null geodesics determined by the above metric.
	Null geodesic equations in a static, spherically symmetric spacetime possess conserved quantities corresponding to the energy and the angular momentum of the photons.
	Due to the conservation of the angular momentum, it is sufficient to consider trajectories on the equatorial plane $\theta=\pi/2$ without loss of generality.
	Then, the geodesic equations reduce to 
	\begin{equation}
		\epsilon=\left(1-\frac{2M}{r}+\frac{9M^2}{8r^2}\right)\frac{\mathrm{d}t}{\mathrm{d}\tau},\ \ L=r^2\frac{\mathrm{d}\phi}{\mathrm{d}\tau},
	\end{equation}
	with the constraint $\mathrm{d}s^2=0$, where $\tau$ is an affine parameter of the trajectory and $\epsilon$ and $L$ are constants.
	Combining these equations to eliminate $t$ and $\tau$, we obtain
	\begin{equation}
		\left(\frac{\mathrm{d}\phi}{\mathrm{d}r}\right)^2=\frac{L^2/r^4}{\epsilon^2-(L^2/r^2)(1-2M/r+9M^2/8r^2)}.
	\end{equation}
	In the following, we assume $L\neq0$, otherwise the photon collides with the singularity without deflection.
	Introducing dimensionless quantities $u=2M/r, x=27M^2\epsilon^2/2L^2$, this equation reduces to
	\begin{equation}
		\left(\frac{\mathrm{d}\phi}{\mathrm{d}u}\right)^2=\frac{1}{8x/27-u^2+u^3-9u^4/32}.\label{dphidu}
	\end{equation}
	Note that the parameter $x$ is related to the impact parameter $b$ as $x=27M^2/2b^2$.

	We focus on trajectories of light coming from and going back to infinity.
	Then, each trajectory has the closest point to the singularity, the radius of which we call $r_0$.
	Note that $u_0=2M/r_0$ is the smallest positive root of the quartic polynomial $8x/27-u^2+u^3-9u^4/32$.
	We observe that $u_0 (>0)$ always exists for arbitrary values of $x (>0)$.
	Then, the deflection angle $\alpha$ as a function of $x$ is given by
	\begin{equation}
		\alpha(x)=2\Delta\phi(x)-\pi,\ \ \Delta\phi(x)=\int_0^{u_0}\frac{\mathrm{d}u}{\sqrt{8x/27-u^2+u^3-9u^4/32}}.\label{integraldphi}
	\end{equation}
	In the following, we also call $\Delta\phi$ the deflection angle for simplicity.

	Although $u_0$ is well-defined for arbitrary values of $x (>0)$, $\Delta\phi(x)$ is divergent at $x=1\ (\Leftrightarrow b^2=27M^2/2)$
	since $u=u_0=3/4$ becomes a triple root of the polynomial in the integrand.
	This divergence physically corresponds to the trajectory of light trapped by the photon sphere.
	More precisely, $x\to1-0$ and $x\to1+0$ correspond to the SDL from outside and inside of the photon sphere respectively
	and the critical impact parameter is given by $b_{\rm m}=3\sqrt{6}M/2$.
	Except for this case, the light ray for a given $x (>0)$ is not trapped and can go back to infinity.
	Therefore we analyze $\Delta\phi(x)$ separately in two segments, i.e. $0<x<1$ and $1<x$.
	Note that the closest point $r_0$ lies outside (inside) the photon sphere for $0<x<1\ (1<x)$.
}}}

\section{Picard-Fuchs equation for $\Delta\phi(x)$ and its solutions\label{sec_PFeq}}
{{{
	We analyze the deflection angle $\Delta\phi(x)$ as a function of $x$ by solving the Picard-Fuchs(PF) equation.
	This method has been used for the cases of the Schwarzschild and RN black hole in \cite{Sasaki_2021,Sasaki_2024},
	and the derivation of the equation is completely the same.
	Therefore, we only show the resultant equation for $\Delta\phi(x)$ below 
	\begin{equation}
		\left[144x(1-x)^2\partial_x^2+144(1-x)(1-2x)\partial_x+27x-31\right]\Delta\phi(x)=\frac{2\sqrt{6}(4-3x)}{\sqrt{x}}. \label{PFeq}
	\end{equation}
	The purpose of this section is to solve this differential equation with the boundary condition shown below on the positive real axis.
	In particular, we derive expressions appropriate around $x=0,1$, and $\infty$, which are regular singularities of (\ref{PFeq})
	as seen from the coefficients of $\partial_x^2$.
	Since the differential operator on the left-hand side of this equation has three regular singularities, 
	this operator can be cast into that of Gauss's hypergeometric equation.
	To see this, we transform the dependent variable as
	\begin{equation}
		\Delta\phi(x)\to f(x)=(1-x)^{1/6}\Delta\phi(x),
	\end{equation}
	which results in 
	\begin{equation}
		\left[x(1-x)\partial_x^2+\left(1-\frac{5x}{3}\right)\partial_x-\frac{7}{144}\right]f(x)=\frac{4-3x}{12\sqrt{6x}(1-x)^{5/6}}.\label{HGeq}
	\end{equation}
	Comparing this equation with the Gauss's hypergeometric equation,
	\begin{equation}
		\left[x(1-x)\partial_x^2+\left(c-(a+b+1)x\right)\partial_x-ab\right]F=0,
	\end{equation}
	which is satisfied by the Gauss's hypergeometric function,
	\begin{equation}
		F={}_2F_1\left[\begin{matrix}a,b\\ c\end{matrix};x\right]:=\sum_{n=0}^\infty\frac{(a)_n(b)_n}{n!(c)_n}x^n,
	\end{equation}
	we see that the equation for $f(x)$ is an inhomogeneous hypergeometric equation with the parameters $a=1/12,b=7/12$, and $c=1$.
	Here, $(a)_n=\Gamma(a+n)/\Gamma(a)$ is the Pochhammer symbol.

}}}

	\subsection{Boundary conditions}
	{{{
		Before solving the PF equation (\ref{PFeq}), we identify the boundary condition for $\Delta\phi$ defined by (\ref{integraldphi}).
		For scattering outside the photon sphere $(0<x<1)$, we consider the limit $x\to+0$, which is called the weak deflection limit (WDL) since 
		this is equivalent to taking the large impact parameter limit $b\to+\infty$ for a fixed background metric.
		In this limit, the endpoint of the integral $u_0$ approaches $0$ so that the interval of the integral shrinks to a single point.
		In order to avoid this and to evaluate the limit, we change the integral variable as $u\to v=\sqrt{27/8x}u$, which results in
		\begin{equation}
			\Delta\phi(x)=\int_0^{v_0}\frac{\mathrm{d}v}{\sqrt{1-v^2+\sqrt{8x/27}v^3-(x/12)v^4}}.
		\end{equation}
		Note that the limit of the endpoint $v_0=\sqrt{27/8x}u_0$ is now kept finite, i.e. $v_0\to1-0$ as $x\to+0$.
		Therefore the WDL of $\Delta\phi(x)$ can be evaluated as 
		\begin{equation}
			\lim_{x\to+0}\Delta\phi(x)=\int_0^1\frac{\mathrm{d}v}{\sqrt{1-v^2}}=\frac{\pi}{2},\label{BC0}
		\end{equation}
		which leads to $\alpha(x)\to0$ as $x\to+0$.
		Similarly, we can derive the following asymptotic form in the limit $x\to+\infty$ for the scattering from inside the photon sphere $(1<x)$:
		\begin{equation}
			\Delta\phi=\frac{3^{1/4}\pi^{3/2}}{2\Gamma^2(\frac{3}{4})}x^{-1/4}+\frac{2\sqrt{2}}{\sqrt{3x}}+\frac{\pi^{3/2}}{3^{1/4}\Gamma^2(\frac{1}{4})}x^{-3/4}+O(x^{-5/4}).\label{BCinfinity}
		\end{equation}
		We call $x\to+\infty$ the head-on limit in the following.
	}}}
	
	\subsection{Solutions for scattering outside the photon sphere: $0<x<1$}
	{{{
		In this subsection, we solve the PF equation (\ref{PFeq}), or equivalently (\ref{HGeq}), in the interval $0<x<1$ with the boundary condition (\ref{BC0}).
		The holomorphic solution $f_1^{(0)}(x)$ to the homogeneous equation for $f(x)$ at $x=0$ is given by Gauss's hypergeometric series,
		\begin{equation}
			f_1^{(0)}(x):={}_2F_1\left[\begin{matrix}\frac{1}{12},\frac{7}{12}\\1\end{matrix};x\right].
		\end{equation}
		Another homogeneous solution $f_2^{(0)}(x)$ at $x=0$ is logarithmically singular and cannot be expressed as a hypergeometric series,
		\begin{equation}
			f_2^{(0)}(x):=\frac{\Gamma(\frac{2}{3})}{\Gamma(\frac{1}{12})\Gamma(\frac{7}{12})}\sum_{n=0}^\infty\frac{(\frac{1}{12})_n(\frac{7}{12})_n}{(n!)^2}\left[2\psi(n+1)-\psi(n+1/12)-\psi(n+7/12)-\log x\right]x^n,
		\end{equation}
		where $\psi(z):=\Gamma'(z)/\Gamma(z)$ is the digamma function.
		Since the inhomogeneous term of (\ref{PFeq}) behaves as $O(x^{1/2})$ when $x\to+0$, we can specify the following inhomogeneous solution:
		\begin{equation}
			\Delta\phi_{\rm I}^{(0)}(x):=\sqrt{\frac{8x}{27}}\sum_{n=0}^\infty I_n^{(0)}x^n.
		\end{equation}
		The coefficients $I_n^{(0)}$ are uniquely determined by (\ref{PFeq}) as
		\begin{equation}
			I_n^{(0)}=\frac{(\frac{2}{3})_n(\frac{4}{3})_n}{(\frac{3}{2})_n(\frac{3}{2})_n}2^n{}_2F_1\left[\begin{matrix}-n,-n-1/2\\-3n-1\end{matrix};\frac{9}{8}\right].\label{In0}
		\end{equation}
		Note that this expression is a special case of the general expression for arbitrary values of $Q^2$ derived in \cite{Sasaki_2024}.
		The boundary condition $\Delta\phi(0)=\pi/2$ then determines the following expression for the deflection angle:
		\begin{equation}
			\Delta\phi(x)=\frac{\pi}{2}(1-x)^{-1/6}f_1^{(0)}(x)+\Delta\phi_{\rm I}^{(0)}(x).\label{dphiWDL}
		\end{equation}
		Since the hypergeometric series is convergent when $|x|<1$, this expression is appropriate in the WDL.

		We next consider the analytic continuation of (\ref{dphiWDL}) to $x\to1-0$.
		Physically, this limit corresponds to the SDL from outside.
		An arbitrary homogeneous solution around $x=1$ for (\ref{HGeq}) can be written as a superposition of the following fundamental solutions:
		\begin{align}
			f_1^{(1)}(x):=&{}_2F_1\left[\begin{matrix}\frac{1}{12},\frac{7}{12}\\\frac{2}{3}\end{matrix};1-x\right],\\
			f_2^{(1)}(x):=&(1-x)^{1/3}{}_2F_1\left[\begin{matrix}\frac{5}{12},\frac{11}{12}\\\frac{4}{3}\end{matrix};1-x\right].
		\end{align}
		The analytic continuation of $f_1^{(0)}(x)$ to $x\to1-0$ is a classical result (see, for example, \cite{HTF1}),
		\begin{equation}
			f_1^{(0)}(x)=
			\frac{\Gamma(\frac{1}{12})\Gamma(\frac{7}{12})}{2\sqrt{3}\pi\Gamma(\frac{2}{3})}f_1^{(1)}(x)
			-\frac{\Gamma(\frac{5}{12})\Gamma(\frac{11}{12})}{2\sqrt{3}\pi\Gamma(\frac{4}{3})}f_2^{(1)}(x) \label{homx=1}
		\end{equation}

		In order to derive a local expression of the inhomogeneous solution $\Delta\phi_{\rm I}^{(0)}$ around $x=1$,
		we express $\Delta\phi_{\rm I}^{(0)}$ in terms of the homogeneous solutions by employing the method of variation of parameters.
		As a result, we obtain
		\begin{equation}
			\Delta\phi_{\rm I}^{(0)}(x)=\frac{\sqrt{6}}{24}\left[\omega_1(x)\int_0^{x}\frac{4-3x'}{\sqrt{x'}(1-x')}\omega_2(x')\mathrm{d}x'
				-\omega_2(x)\int_0^x\frac{4-3x'}{\sqrt{x'}(1-x')}\omega_1(x')\mathrm{d}x'\right], \label{dphi0C}
		\end{equation}
		where $\omega_i(x)=(1-x)^{-1/6}f_i^{(1)}(x)\ (i=1,2)$.
		By expanding $\omega_i(x)$ in powers of $1-x$ and integrating term by term, we can derive a power series expansion of $\Delta\phi_{\rm I}^{(0)}$ at $x=1$.
		In particular, we can show the following asymptotic behavior (see Appendix \ref{Derivation1} for derivation):
		\begin{equation}
			\Delta\phi_{\rm I}^{(0)}(x)=N_1\cdot(1-x)^{-1/6}(1+O(1-x))-N_2\cdot (1-x)^{1/6}(1+O(1-x))+N_{0}\cdot(1+O(1-x)),\ \ x\to1-0, \label{dphi0_continuation}
		\end{equation}
		where $N_1, N_2$, and $N_{0}$ are constants given by
		\begin{align}
			N_1=&\frac{7\sqrt{\pi}\Gamma(\frac{1}{6})}{16\sqrt{6}\Gamma(\frac{2}{3})}
				{}_4F_3\left[\begin{matrix}\frac{5}{12},\frac{11}{12},\frac{1}{6},\frac{31}{24}\\\frac{4}{3},\frac{5}{3},\frac{7}{24}\end{matrix};1\right],\label{N1}\\
			N_2=&\frac{\sqrt{6}\pi^{3/2}}{4\Gamma(\frac{1}{3})\Gamma(\frac{1}{6})}
				{}_4F_3\left[\begin{matrix}\frac{1}{12},\frac{7}{12},-\frac{1}{6},\frac{23}{24}\\\frac{2}{3},\frac{4}{3},-\frac{1}{24}\end{matrix};1\right],\label{N2}\\
			N_0=&-\frac{\sqrt{6}}{2}.
		\end{align}
		Since $\Delta\phi_{\rm I}^{(0)}$ is a solution to (\ref{PFeq}) and the inhomogeneous term of (\ref{PFeq}) is holomorphic at $x=1$,
		the power series associated with the constants $N_1$ and $N_2$ in the above expression must be homogeneous solutions 
		while that for $N_0$ corresponds to the specific inhomogeneous solution $\Delta\phi_{\rm I}^{(1)}$ that is holomorphic at $x=1$.
		Thus, we conclude
		\begin{equation}
			\Delta\phi_{\rm I}^{(0)}(x)=N_1\omega_1(x)-N_2\omega_2(x)+\Delta\phi_{\rm I}^{(1)}(x).\label{inhomx=1}
		\end{equation}
		The specific inhomogeneous solution $\Delta\phi_{\rm I}^{(1)}(x)$ is defined by the following power series:
		\begin{equation}
			\Delta\phi_{\rm I}^{(1)}(x):=\sum_{n=0}^\infty I_n^{(1)}(1-x)^n,
		\end{equation}
		where the coefficients $I_n^{(1)}$ are uniquely determined by the recurrence relation derived from (\ref{PFeq}),
		\begin{equation}
			I_0^{(1)}=-\frac{\sqrt{6}}{2},\ \ \left(n^2-\frac{1}{36}\right)I_n^{(1)}-\left(n-\frac{1}{4}\right)\left(n-\frac{3}{4}\right)I_{n-1}^{(1)}=\frac{\sqrt{6}}{72}\frac{(-\frac{1}{2})_n(\frac{7}{8})_n}{n!(-\frac{1}{8})_n},\ (n=1,2,3,\cdots).\label{In1}
		\end{equation}

		Combining the continuation formulae for the homogeneous solution (\ref{homx=1}) and the inhomogeneous solution (\ref{inhomx=1}), 
		we obtain the following local expression of the deflection angle valid for $x\to1-0$:
		\begin{equation}
			\Delta\phi(x)=\left(\frac{\Gamma(\frac{1}{12})\Gamma(\frac{7}{12})}{4\sqrt{3}\Gamma(\frac{2}{3})}+N_1\right)\omega_1(x)
			-\left(\frac{\Gamma(\frac{5}{12})\Gamma(\frac{11}{12})}{4\sqrt{3}\Gamma(\frac{4}{3})}+N_2\right)\omega_2(x)+\Delta\phi_{\rm I}^{(1)}(x).
		\end{equation}
		We comment on the value of the constants $N_1$ and $N_2$.
		By numerically evaluating these constants, we found the constants in each of the parenthesis in the above expression
		are equal up to (at least) 20 digits, namely
		\begin{align}
			N_1=&\frac{\Gamma(\frac{1}{12})\Gamma(\frac{7}{12})}{4\sqrt{3}\Gamma(\frac{2}{3})}=1.8738033603506541537\cdots,\\
			N_2=&\frac{\Gamma(\frac{5}{12})\Gamma(\frac{11}{12})}{4\sqrt{3}\Gamma(\frac{4}{3})}=0.36299153688172872301\cdots.
		\end{align}
		Although we couldn't prove this, we assume this holds exactly.
		As a result, we have
		\begin{equation}
			\Delta\phi(x)=\frac{\Gamma(\frac{1}{12})\Gamma(\frac{7}{12})}{2\sqrt{3}\Gamma(\frac{2}{3})}\omega_1(x)
			-\frac{\Gamma(\frac{5}{12})\Gamma(\frac{11}{12})}{2\sqrt{3}\Gamma(\frac{4}{3})}\omega_2(x)+\Delta\phi_{\rm I}^{(1)}(x).\label{dphiSDLoutside}
		\end{equation}

	}}}

	\subsection{Solutions for scattering inside the photon sphere: $1<x$}
	{{{
		In this subsection, we solve the PF equation (\ref{PFeq}) (or (\ref{HGeq})) in the region $1<x$ with the boundary condition (\ref{BCinfinity}).
		The fundamental homogeneous solutions to (\ref{HGeq}) are given by
		\begin{align}
			f_1^{(\infty)}(x):=&x^{-1/12}{}_2F_1\left[\begin{matrix}\frac{1}{12},\frac{1}{12}\\\frac{1}{2}\end{matrix};\frac{1}{x}\right],\\
			f_2^{(\infty)}(x):=&x^{-7/12}{}_2F_1\left[\begin{matrix}\frac{7}{12},\frac{7}{12}\\\frac{3}{2}\end{matrix};\frac{1}{x}\right].
		\end{align}
		Using these hypergeometric series, we define $\omega_i^{(\infty)}(x)=(x-1)^{-1/6}f_i^{(\infty)}(x)\ (i=1,2)$ as homogeneous solutions to (\ref{PFeq}).
		On the other hand, we have the following inhomogeneous solution to (\ref{PFeq}):
		\begin{equation}
			\Delta\phi_{\rm I}^{(\infty)}:=\frac{1}{\sqrt{x}}\sum_{n=0}^\infty I_n^{(\infty)}x^{-n},
		\end{equation}
		where the coefficients $I_n^{(\infty)}$ are given by
		\begin{equation}
			I_n^{(\infty)}=\sqrt{\frac{8}{3}}(-3)^n{}_2F_1\left[\begin{matrix}-2n,n+1\\\frac{3}{2}\end{matrix};\frac{8}{9}\right].\label{Ininf}
		\end{equation}
		Taking the boundary condition at infinity (\ref{BCinfinity}) into account, we obtain
		\begin{equation}
			\Delta\phi(x)=\frac{3^{1/4}\pi^{3/2}}{2\Gamma^2(\frac{3}{4})}(x-1)^{-1/6}f_1^{(\infty)}(x)
			+\frac{\pi^{3/2}}{3^{1/4}\Gamma^2(\frac{1}{4})}(x-1)^{-1/6}f_2^{(\infty)}(x)+\Delta\phi_{\rm I}^{(\infty)}(x).\label{dphiinfinity}
		\end{equation}

		We next consider the analytic continuation of (\ref{dphiinfinity}) to $x\to1+0$ as in the previous subsection.
		For the homogeneous solutions $f_{1,2}^{(\infty)}(x)$, the continuation formulae are as follows\cite{HTF1}:
		\begin{align}
			f_1^{(\infty)}(x)&=\frac{2\pi^{3/2}}{\sqrt{3}\Gamma^2(\frac{5}{12})\Gamma(\frac{2}{3})}\tilde{f}_1^{(1)}(x)-\frac{2\pi^{3/2}}{\sqrt{3}\Gamma^2(\frac{1}{12})\Gamma(\frac{4}{3})}\tilde{f}_2^{(1)}(x),\label{infto1_1}\\
			f_2^{(\infty)}(x)&=\frac{\pi^{3/2}}{\sqrt{3}\Gamma^2(\frac{11}{12})\Gamma(\frac{2}{3})}\tilde{f}_1^{(1)}(x)-\frac{\pi^{3/2}}{\sqrt{3}\Gamma^2(\frac{7}{12})\Gamma(\frac{4}{3})}\tilde{f}_2^{(1)}(x),\label{infto1_2}
		\end{align}
		where
		\begin{align}
			\tilde{f}_1^{(1)}(x)&:=x^{-1/12}{}_2F_1\left[\begin{matrix}\frac{1}{12},\frac{1}{12}\\\frac{2}{3}\end{matrix};1-\frac{1}{x}\right],\\
				\tilde{f}_2^{(1)}(x)&:=x^{-5/12}(x-1)^{1/3}{}_2F_1\left[\begin{matrix}\frac{5}{12},\frac{5}{12}\\\frac{4}{3}\end{matrix};1-\frac{1}{x}\right].
		\end{align}
		is another set of fundamental solutions at $x=1$.
		In order to continue the inhomogeneous solution $\Delta\phi_{\rm I}^{(\infty)}(x)$ to $x\to1+0$, 
		we again express $\Delta\phi_{\rm I}^{(\infty)}(x)$ in terms of the homogeneous solutions $\omega_i^{(\infty)}\ (i=1,2)$.
		The method of variation of parameters leads to the following expression:
		\begin{equation}
			\Delta\phi_{\rm I}^{(\infty)}(x)=C_1(x)\omega_1^{(\infty)}(x)+C_2(x)\omega_2^{(\infty)}(x),\label{dphiinfC}
		\end{equation}
		where $C_1(x)$ and $C_2(x)$ obey
		\begin{equation}
			\partial_xC_1(x)=-\frac{\sqrt{6}}{36}\frac{3x-4}{(x-1)\sqrt{x}}\omega_2^{(\infty)}(x),\ \ 
			\partial_xC_2(x)=\frac{\sqrt{6}}{36}\frac{3x-4}{(x-1)\sqrt{x}}\omega_1^{(\infty)}(x).\label{Ceqinf}
		\end{equation}
		Integration constants for $C_{1,2}(x)$ must be set to match the asymptotic behavior $\Delta\phi_{\rm I}^{(\infty)}(x)=I_0^{(\infty)}x^{-1/2}+I_1^{(\infty)}x^{-3/2}+\cdots$.
		This boundary condition can be realized by expanding the integrand in terms of $1/x$ and integrating term by term.
		After that, we change the variable of expansion to $x-1$, which results in (see Appendix \ref{Derivation2} for derivation)
		\begin{equation}
			\Delta\phi_{\rm I}^{(\infty)}(x)=\frac{\sqrt{6}}{24}\tilde{N}_2(x-1)^{-1/6}(1+O(x-1))-\frac{\sqrt{6}}{24}\tilde{N}_1(x-1)^{1/6}(1+O(x-1))-\frac{\sqrt{6}}{2}(1+O(x-1)),\ (x\to1+0) \label{dphiinf_continuation}
		\end{equation}
		where the constants $\tilde{N}_i\ (i=1,2)$ are given by (\ref{Nt1}) and (\ref{Nt2}).
		Combining the continuation formulae for the homogeneous solutions (\ref{infto1_1}) and (\ref{infto1_2}),
		and that for the inhomogeneous solution (\ref{dphiinf_continuation}), we obtain the following asymptotic form for the deflection angle $\Delta\phi$:
		\begin{equation}
			\Delta\phi=\left(\frac{\Gamma(\frac{1}{12})\Gamma(\frac{7}{12})}{4\Gamma(\frac{2}{3})}+\frac{\sqrt{6}}{24}\tilde{N}_2\right)(x-1)^{-1/6}
			-\left(\frac{\Gamma(\frac{5}{12})\Gamma(\frac{11}{12})}{4\Gamma(\frac{4}{3})}+\frac{\sqrt{6}}{24}\tilde{N}_1\right)(x-1)^{1/6}-\frac{\sqrt{6}}{2}+\cdots.
		\end{equation}
		Interestingly, by numerically evaluating the constants $\tilde{N}_1$ and $\tilde{N}_2$, the constants in each of the parenthesis in this expression are found to be identical up to (at least) 20 digits,
		\begin{align}
			\frac{\Gamma(\frac{1}{12})\Gamma(\frac{7}{12})}{4\Gamma(\frac{2}{3})}=\frac{\sqrt{6}}{24}\tilde{N}_2=3.2455226235206265142\cdots,\\
			\frac{\Gamma(\frac{5}{12})\Gamma(\frac{11}{12})}{4\Gamma(\frac{4}{3})}=\frac{\sqrt{6}}{24}\tilde{N}_1=0.62871978459666614131\cdots.
		\end{align}
		We again assume these equalities hold exactly, which finally leads to the following expression:
		\begin{equation}
			\Delta\phi=\frac{\Gamma(\frac{1}{12})\Gamma(\frac{7}{12})}{2\Gamma(\frac{2}{3})}\tilde{\omega}_1(x)-\frac{\Gamma(\frac{5}{12})\Gamma(\frac{11}{12})}{2\Gamma(\frac{4}{3})}\tilde{\omega}_2(x)+\Delta\phi_{\rm I}^{(1)}(x),\label{dphiSDLinside}
		\end{equation}
		where $\tilde{\omega}_i(x)=(x-1)^{-1/6}\tilde{f}_i^{(1)}(x)\ (i=1,2)$.
	}}}

\section{Accuracy of the series expansions\label{sec_accuracy}}
{{{
	In the previous section, we have derived four expressions for the deflection angle $\Delta\phi$, (\ref{dphiWDL}), (\ref{dphiSDLoutside}), (\ref{dphiSDLinside}), and (\ref{dphiinfinity}), which are valid around $x=0, x=1-0, x=1+0$, and $x=\infty$ respectively.
	Note that each expression consists of a sum of the homogeneous solutions and the specific inhomogeneous solution to the PF equation (\ref{PFeq}).
	Since the homogeneous solutions are given by the Gauss's hypergeometric series multiplied by the powers of $x$ and/or $x-1$,
	one can easily derive explicit expressions up to the desired orders.
	For the inhomogeneous solutions at $x=0$ and $x=\infty$, namely $\Delta\phi_{\rm I}^{(0)}(x)$ and $\Delta\phi_{\rm I}^{(\infty)}(x)$, 
	we showed the explicit expressions for the expansion coefficients in (\ref{In0}) and (\ref{Ininf}) respectively.
	For the inhomogeneous solution $\Delta\phi_{\rm I}^{(1)}(x)$ at $x=1$, we gave the recurrence relation satisfied by the expansion coefficient in (\ref{In1}).
	Thus, the results so far enable us to approximate the deflection angle $\Delta\phi$ around each singularity with desired accuracy by taking sufficiently higher-order terms into account.
	In this section, we compare the analytical results so far with the numerical integration of $\Delta\phi$.

	In Figures \ref{fig_dphi} and \ref{fig_error}, we plotted $\Delta\phi$ as a function of $b/b_{\rm m}=1/\sqrt{x}$ and the relative error of the analytic approximations
	at each singularity with the numerical integration respectively.
	These figures show that our formulae give correct asymptotic approximations as the relative errors converge to $0$ at each singularity
	and that the inclusion of the higher order terms makes the approximations more accurate.
	In the following, we focus on the SDL.

	For the SDL from the outside $(x\to1-0 \Leftrightarrow b/b_{\rm m}\to1+0)$, we also plotted the asymptotic formula derived by Tsukamoto in \cite{Tsukamoto2020}.
	Tsukamoto derived an asymptotic formula for the total deflection angle $\alpha=2\Delta\phi-\pi$ (see (\ref{SDLpower}) and (\ref{cf_Tsukamoto2020}))
	by directly manipulating the integral expression in the limit $b/b_{\rm m}\to1+0$, which is equivalent to 
	\begin{equation}
		\Delta\phi_{\rm Tsukamoto}=\frac{\bar{c}/2}{(b/b_{\rm m}-1)^{1/6}}+\frac{\bar{d}+\pi}{2}+O((b/b_{\rm m}-1)^{1/6}),\ \ \bar{c}=2^{5/3}\cdot 3^{1/2},\ \ \bar{d}=-\sqrt{6}-\pi. \label{SDLoutTsukamoto}
	\end{equation}
	On the other hand, by truncating the terms of order $O((1-x)^{1/6})$ and higher in (\ref{dphiSDLoutside}) and approximating as $1-x\simeq2(b/b_{\rm m}-1)$, 
	we obtain
	\begin{equation}
		\Delta\phi=\frac{\Gamma(\frac{1}{12})\Gamma(\frac{7}{12})}{2^{7/6}\sqrt{3}\Gamma(\frac{2}{3})}\frac{1}{(b/b_{\rm m}-1)^{1/6}}-\frac{\sqrt{6}}{2}+O((b/b_{\rm m}-1)^{1/6}). \label{SDLoutO(1)}
	\end{equation}
	Although the values of the $O(1)$-terms in (\ref{SDLoutTsukamoto}) and (\ref{SDLoutO(1)}) are consistent,
	the coefficients of the divergent terms are different,
	\begin{equation}
		\frac{\bar{c}}{2}\simeq2.749,\ \ \frac{\Gamma(\frac{1}{12})\Gamma(\frac{7}{12})}{2^{7/6}\sqrt{3}\Gamma(\frac{2}{3})}\simeq3.339.
	\end{equation}
	In contrast to the relative error of eq.(\ref{SDLoutO(1)}), that of eq.(\ref{SDLoutTsukamoto}) doesn't seem to converge to $0$ in the limit $b/b_{\rm m}\to1+0$ as shown in the inset of Fig.\ref{fig_error} (b).
	Thus, we conclude that the asymptotic formula (\ref{SDLoutTsukamoto}) (or (\ref{SDLpower}) with (\ref{cf_Tsukamoto2020})) derived in \cite{Tsukamoto2020} is incorrect.

	Although (\ref{SDLoutO(1)}) gives a correct asymptotic approximation in the limit $b/b_{\rm m}\to1+0$,
	the relative error exceeds $10\%$ for $b/b_{\rm m}\gtrsim1.03\ \Leftrightarrow \alpha\lesssim5.46<2\pi$.
	On the other hand, the $(n=0)$-formula, which includes not only the divergent and $O(1)$-terms but also the term of order $O((1-x)^{1/6})$,
	\begin{equation}
		\Delta\phi(x)=\frac{\Gamma(\frac{1}{12})\Gamma(\frac{7}{12})}{2\sqrt{3}\Gamma(\frac{2}{3})}(1-x)^{-1/6}
		-\frac{\Gamma(\frac{5}{12})\Gamma(\frac{11}{12})}{2\sqrt{3}\Gamma(\frac{4}{3})}(1-x)^{1/6}-\frac{\sqrt{6}}{2}+O((1-x)^{5/6}),
	\end{equation}
	is much more accurate as the relative error is less than $1\%$ for $b/b_{\rm m}\lesssim1.77\ \Leftrightarrow\ x\gtrsim0.32$, 
	which corresponds to $\alpha\gtrsim1.00\sim\pi/3$.
	Usually, the deflection angle in the SDL is well approximated only by logarithmically divergent term and $O(1)$-term\cite{Bozza_2002,Tsukamoto_2017}.
	For example, the relative error of this approximation for $\alpha$ in the Schwarzschild spacetime from the exact value calculated numerically 
	is less than $1\%$ when $\alpha\gtrsim 2\pi$\cite{Bozza_2001,Bozza_2002}.
	Therefore $O((b/b_{\rm m}-1)^{1/6})$-term should be included to calculate with the same level of accuracy as the standard logarithmic formula in the SDL from the outside.

	The accuracy of the formula of the same order is worse in the SDL from inside the photon sphere, i.e. $x\to1+0$.
	For example, the $(n=0)$-formula,
	\begin{equation}
		\Delta\phi(x)=\frac{\Gamma(\frac{1}{12})\Gamma(\frac{7}{12})}{2\Gamma(\frac{2}{3})}\left(1-\frac{1}{x}\right)^{-1/6}
		-\frac{\Gamma(\frac{5}{12})\Gamma(\frac{11}{12})}{2\Gamma(\frac{4}{3})}\left(1-\frac{1}{x}\right)^{1/6}-\frac{\sqrt{6}}{2}+O((1-1/x)^{5/6}),\label{n=0SDLinside}
	\end{equation}
	yields a relative error of more than $10\%$ when $\alpha\sim2\pi$ and
	we have to include up to $O((1-1/x)^{13/6})$-term to accomplish an accuracy of less than $1\%$ there (see Appendix \ref{appendix_coeff} for the explicit expression).

	\begin{figure}[htb]
		\centering
		\begin{minipage}{0.49\linewidth}
			\centering
			\includegraphics[keepaspectratio,width=0.9\linewidth]{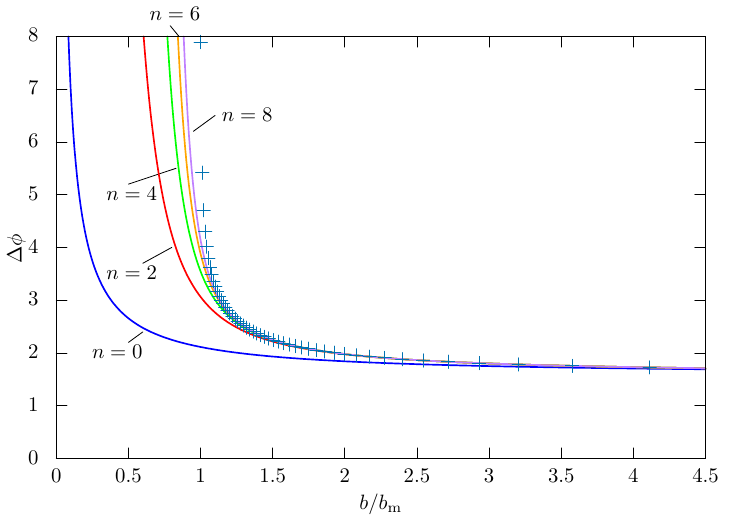}
			\subcaption{WDL $(x\to+0 \Leftrightarrow b/b_{\rm m}\to\infty)$}
			\label{fig:WDL}
		\end{minipage}
		\begin{minipage}{0.49\linewidth}
			\centering
			\includegraphics[keepaspectratio,width=0.9\linewidth]{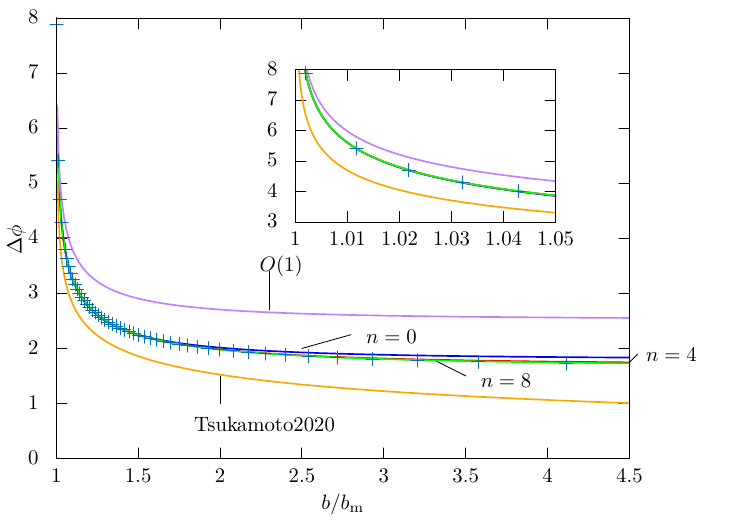}
			\subcaption{SDL from outside $(x\to1-0 \Leftrightarrow b/b_{\rm m}\to1+0)$}
			\label{fig:SDLout}
		\end{minipage}
		\begin{minipage}{0.49\linewidth}
			\centering
			\includegraphics[keepaspectratio,width=0.9\linewidth]{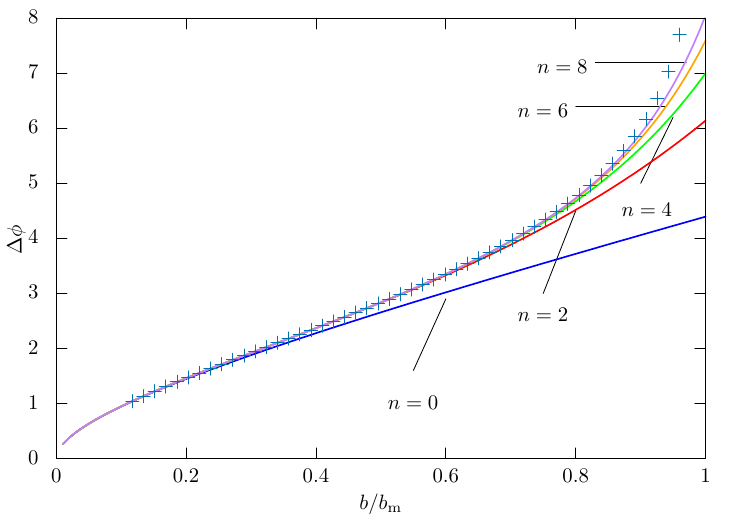}
			\subcaption{Head-on limit $(x\to\infty \Leftrightarrow b/b_{\rm m}\to+0)$}
			\label{fig:HeadOn}
		\end{minipage}
		\begin{minipage}{0.49\linewidth}
			\centering
			\includegraphics[keepaspectratio,width=0.9\linewidth]{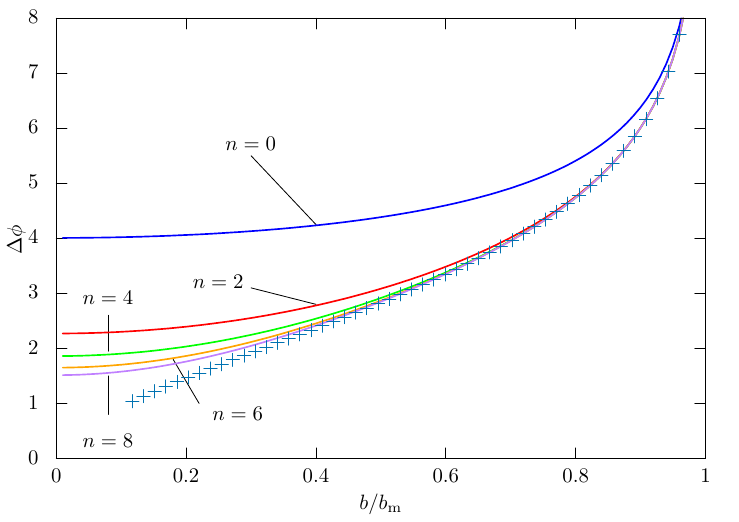}
			\subcaption{SDL from inside $(x\to1+0 \Leftrightarrow b/b_{\rm m}\to1-0)$}
			\label{fig:SDLin}
		\end{minipage}
		\caption{Comparison of the analytic expressions (solid curves) with the numerical integration of (\ref{integraldphi}) (cross marks).
		In either case, the value of $n$ represents the maximum order of the power series (see Appendix \ref{appendix_coeff} for the explicit coefficients).
		In (b), we plotted the asymptotic formula (\ref{SDLoutTsukamoto}) and the corresponding formula (\ref{SDLoutO(1)}) truncated up to the $O(1)$-term as well,
		which are denoted by Tsukamoto2020 and $O(1)$ in the figure.}
		\label{fig_dphi}
	\end{figure}

	\begin{figure}[htb]
		\centering
		\begin{minipage}{0.49\linewidth}
			\centering
			\includegraphics[keepaspectratio,width=0.9\linewidth]{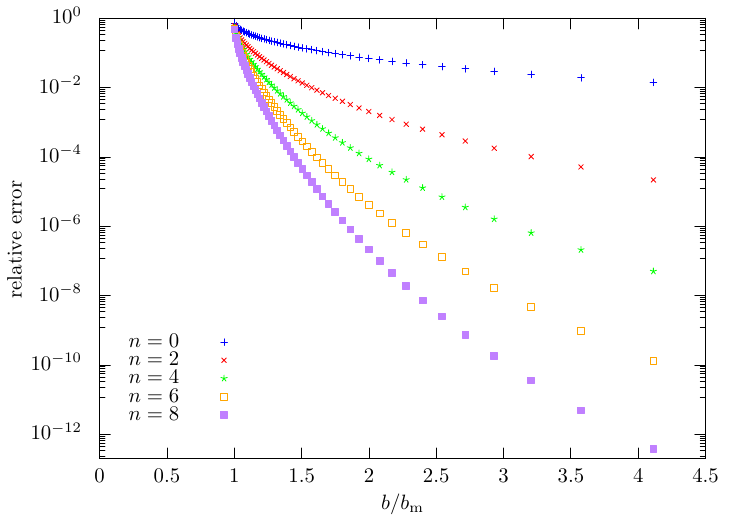}
			\subcaption{WDL $(x\to+0 \Leftrightarrow b/b_{\rm m}\to\infty)$}
			\label{fig:WDL_error}
		\end{minipage}
		\begin{minipage}{0.49\linewidth}
			\centering
			\includegraphics[keepaspectratio,width=0.9\linewidth]{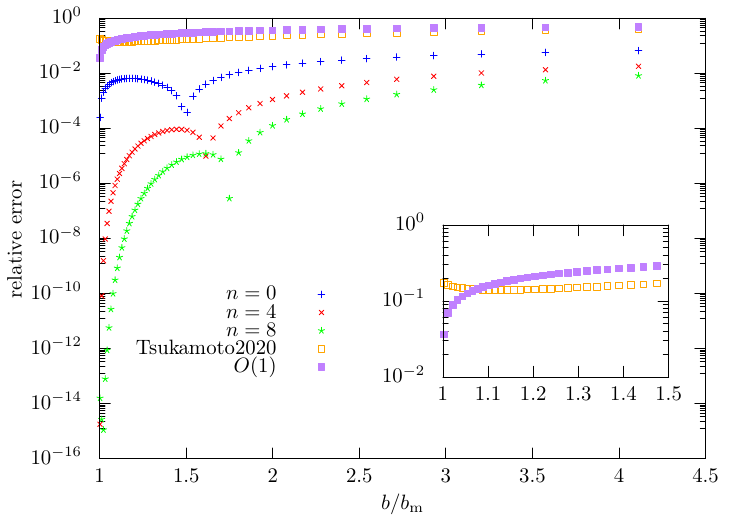}
			\subcaption{SDL from outside $(x\to1-0 \Leftrightarrow b/b_{\rm m}\to1+0)$}
			\label{fig:SDLout_error}
		\end{minipage}
		\begin{minipage}{0.49\linewidth}
			\centering
			\includegraphics[keepaspectratio,width=0.9\linewidth]{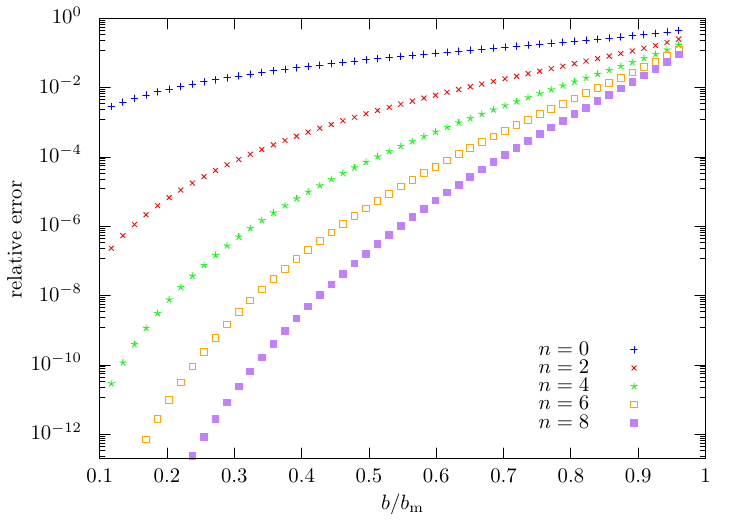}
			\subcaption{Head-on limit $(x\to\infty \Leftrightarrow b/b_{\rm m}\to+0)$}
			\label{fig:HeadOn_error}
		\end{minipage}
		\begin{minipage}{0.49\linewidth}
			\centering
			\includegraphics[keepaspectratio,width=0.9\linewidth]{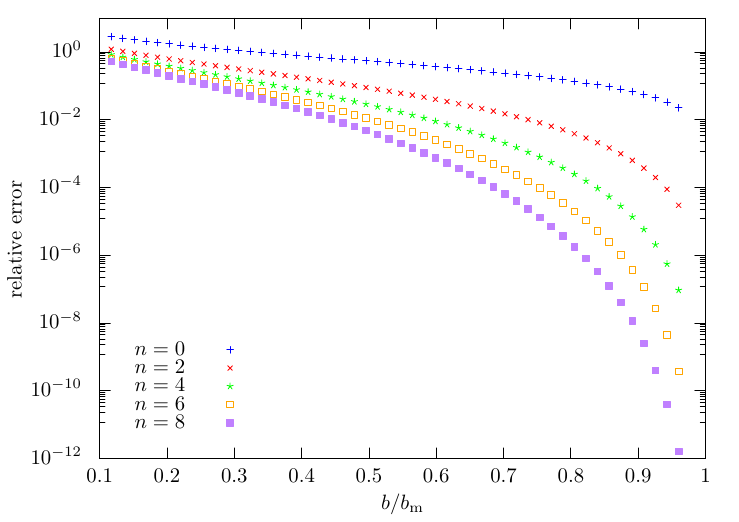}
			\subcaption{SDL from inside $(x\to1+0 \Leftrightarrow b/b_{\rm m}\to1-0)$}
			\label{fig:SDLin_error}
		\end{minipage}
		\caption{Relative errors of the analytic expressions compared with the numerical integration of (\ref{integraldphi}).
		In either case, the value of $n$ represents the maximum order of the power series (see Appendix \ref{appendix_coeff} for the explicit coefficients).
		In (b), we plotted the asymptotic formula (\ref{SDLoutTsukamoto}) and the corresponding formula (\ref{SDLoutO(1)}) truncated up to the $O(1)$-term as well,
		which are enlarged in the inset.
		These are denoted by Tsukamoto2020 and $O(1)$ in the figure.}
		\label{fig_error}
	\end{figure}
}}}

\section{Lens equation and Image position\label{sec_lenseq}}
{{{
	After examining the accuracy of calculating the deflection angle, we consider the angular positions of images as an observable.
	We begin with the exact lens equation\cite{FrittelliNewman_1999,FrittelliKlingNewman_2000,Perlick2004,Bozza_2008b}.
	In the following, we assume that the closest point exists on the trajectory between the source and the observer.
	Extension of the following arguments to the case where this assumption doesn't hold is straightforward.
	In this case, by integrating the geodesic equation (\ref{dphidu}), we obtain the following lens equation:
	\begin{equation}
		\phi=\left(\int_{u_{\rm O}}^{u_0}+\int_{u_{\rm S}}^{u_0}\right)\frac{\mathrm{d}u}{\sqrt{8x/27-u^2+u^3-9u^4/32}},
	\end{equation}
	where $\phi$ represents the difference of the azimuthal angle between the source and the observer (see Figure \ref{fig:configuration}).
	Note that $\phi$ can be indefinitely large in this equation when the geodesic turns around the singularity many times.
	If one would like to consider a single-valued angle $\phi_{\rm eff}$ which is restricted to, for example, $0\leq\phi_{\rm eff}<2\pi$, 
	they are related by $\phi=\phi_{\rm eff}+2\pi N$ for some integer $N$.
	Mathematically, we see this equation to be solved for $x$ (impact parameter) for given $u_{\rm O,S}$ and $\phi$ (or $\phi_{\rm eff}$ and $N$).
	Although the source angle seen by the observer is not well-defined except when $u_{\rm O,S}\ll1$ (see below), 
	perfectly aligned configurations $\phi=\pi,3\pi,5\pi,\cdots$ are well-defined for $0<u_{\rm O,S}<u_0$ due to the spherical symmetry.

	The right-hand side of the lens equation can be cast into the following form:
	\begin{equation}
		\phi=2\Delta\phi(x)-\left(\int_0^{u_{\rm O}}+\int_0^{u_{\rm S}}\right)\frac{\mathrm{d}u}{\sqrt{8x/27-u^2+u^3-9u^4/32}}. \label{lenseq_exact}
	\end{equation}
	This expression is appropriate to consider the limit where the source and/or the observer lie far away from the lens, i.e. $u_{\rm S}\ll1$ and/or $u_{\rm O}\ll1$.
	In this paper, we assume both $u_{\rm O,S}\ll1$ hold and ignore the cubic and quartic terms in the integrand, which leads to the following approximate lens equation:
	\begin{equation}
		\phi=2\Delta\phi(x)-\arcsin\left(\sqrt{\frac{27}{8x}}u_{\rm O}\right)-\arcsin\left(\sqrt{\frac{27}{8x}}u_{\rm S}\right).\label{Lenseq_far}
	\end{equation}

	In order to rewrite this equation in terms of the variables used in the gravitational lensing problem,
	we first recall the definitions $u=2M/r$ and $x=27M^2/2b^2$.
	Then, we see that the second term in the right-hand side is nothing but the image position $\Theta=\arcsin(b/r_{\rm O})$.
	Usually, lens equations are written in terms of the distances among the image, lens, and observer planes.
	These are related to the radial coordinate distances $r_{\rm O,S}$ from the lens by 
	\begin{equation}
		D_{\rm OL}=r_{\rm O},\ \ D_{\rm LS}=r_{\rm S}|\cos(\pi-\phi)|.
	\end{equation}
	Note that $\cos(\pi-\phi)$ becomes negative in the retro-lensing case.
	With the assumption $u_{\rm O,S}\ll1$, we can introduce the source position $\beta$ seen by the observer as
	\begin{equation}
		D_{\rm OS}\tan\beta=D_{\rm LS}|\tan(\pi-\phi)|.
	\end{equation}
	If we restrict the problem to the standard lens configuration, namely $0\leq\beta<\pi/2$ and $2N\pi\leq\pi-\phi<(2N+1/2)\pi$ for some integer $N$,
	we can solve as 
	\begin{equation}
		\pi-\phi=\arctan\left(\frac{D_{\rm OS}}{D_{\rm LS}}\tan\beta\right),\ \ {\rm mod}\ 2\pi.
	\end{equation}
	Combining the discussion so far, the lens equation (\ref{Lenseq_far}) becomes
	\begin{equation}
		\alpha-\Theta-\arcsin\left(\frac{D_{\rm OL}\sin\Theta}{\sqrt{(D_{\rm LS})^2+(D_{\rm OS})^2\tan^2\beta}}\right)
		+\arctan\left(\frac{D_{\rm OS}}{D_{\rm LS}}\tan\beta\right)=0,\ \ {\rm mod}\ 2\pi, \label{Takizawa_eq}
	\end{equation}
	which is equivalent to the lens equation derived in \cite{Takizawa2020}.
	We also note that this equation is equivalent to the Ohanian-Bozza (OB) equation\cite{Ohanian_1987,Bozza_2008b}.

	\begin{figure}[htb]
		\centering
		\includegraphics[keepaspectratio,width=0.9\linewidth]{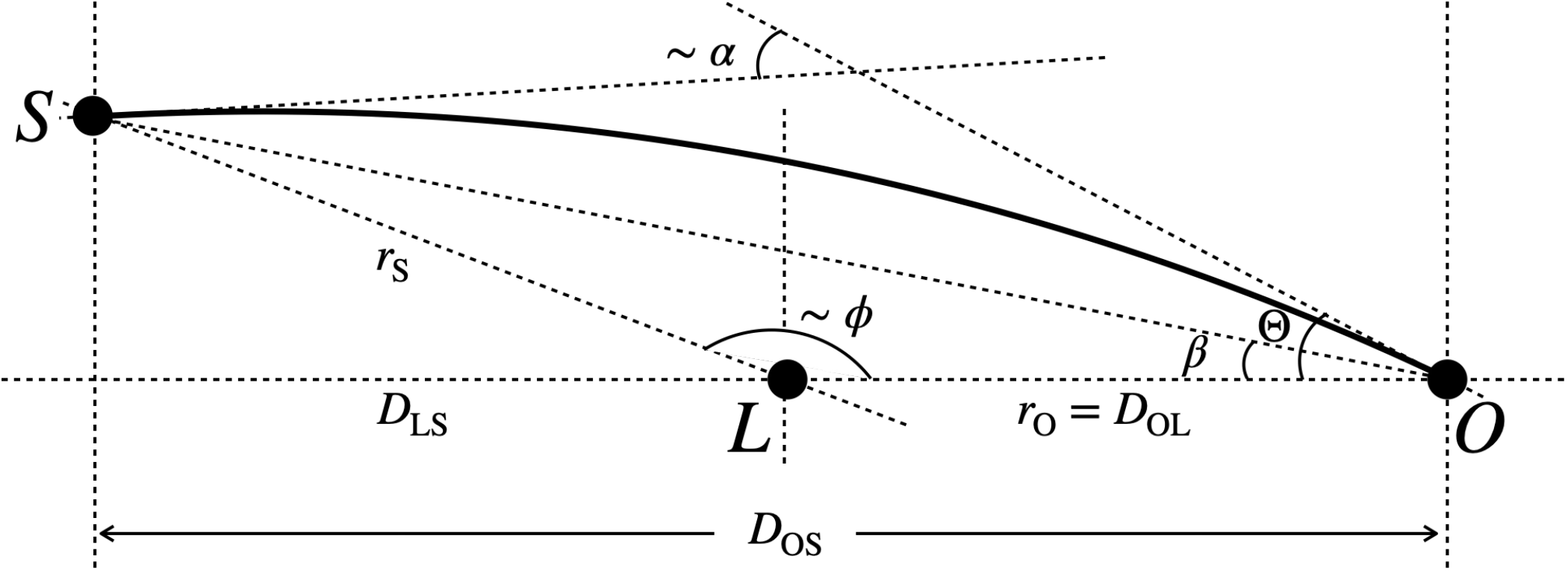}
		\caption{
			Lens configuration. 
		}
		\label{fig:configuration}
	\end{figure}

	Instead of the known form (\ref{Takizawa_eq}), we go back to (\ref{Lenseq_far}) and solve it for $x$.
	Since this equation is transcendental, we expand the right-hand side with respect to $x$ and solve order-by-order.
	We here consider the SDL and expand in powers of $1-x$ or $1-1/x$ in accord with $x\to1-0$ or $x\to1+0$.

	For the SDL from outside $(x\to1-0)$, the lens equation (\ref{Lenseq_far}) is expanded as
	\begin{align}
		\phi=&(1-x)^{-1/6}\left[a_0+a_1(1-x)+\cdots\right]-(1-x)^{1/6}\left[b_0+b_1(1-x)+\cdots\right] \notag\\
		&\hspace{4em}-\left[c_0+c_1(1-x)+\cdots\right], \label{lenseq_SDLout}
	\end{align}
	where the coefficients $a_n,b_n$, and $c_n$ are determined by (\ref{dphiSDLoutside}) and the derivatives of $\arcsin(\sqrt{27/8x}u)$.
	In particular, $a_0=\Gamma(1/12)\Gamma(7/12)/\sqrt{3}\Gamma(2/3), b_0=\Gamma(5/12)\Gamma(11/12)/\sqrt{3}\Gamma(4/3)$, 
	and $c_0=\sqrt{6}+\arcsin(\sqrt{27/8}u_{\rm O})+\arcsin(\sqrt{27/8}u_{\rm S})$.
	One can, at least formally, solve for $1-x$ as a power series in terms of $1/\phi$,
	\begin{equation}
		1-x=\left(\frac{a_0}{\phi}\right)^6\left(1+\frac{A_1}{\phi}+\frac{A_2}{\phi^2}+\cdots\right),
	\end{equation}
	which is expected to converge in the limit $\phi\to+\infty$.
	Instead of doing such an analysis, we approximately solve (\ref{lenseq_SDLout}) by truncating the power series at a finite order.
	In the literature, the deflection angle has been approximated by the leading divergent term and the next-leading $O(1)$-term\cite{Tsukamoto2020b,Tsukamoto2020,Tsukamoto2021},
	which corresponds to taking the terms including $a_0$ and $c_0$ into account,
	\begin{equation}
		\phi=a_0(1-x)^{-1/6}-c_0.
	\end{equation}
	In this approximation, we have
	\begin{equation}
		x=1-\left(\frac{a_0}{\phi+c_0}\right)^6. \label{xO(1)}
	\end{equation}
	Similarly, if we include the next order, i.e. $O((1-x)^{1/6})$-term, we obtain 
	\footnote{In deriving this expression, we have discarded the other solution to the quadratic equation for $(1-x)^{-1/6}$
	taking the boundary condition $x\to1-0$ as $\phi\to+\infty$ into account.}
	\begin{equation}
		x=1-\left(\frac{2a_0}{\phi+c_0}\right)^6\left(1+\sqrt{1+\frac{4a_0b_0}{(\phi+c_0)^2}}\right)^{-6}. \label{xSDL}
	\end{equation}
	By inserting these approximate solutions into $\Theta=\arcsin(\sqrt{27/8x}u_{\rm O})\simeq\sqrt{27/8x}u_{\rm O}$,
	we obtain the image position $\Theta$ as a function of the lens configuration $\phi$ and $u_{\rm O,S}$.
	In the following discussion, the approximate solutions (\ref{xO(1)}) and (\ref{xSDL}) are denoted by $x_{O(1)}$ and $x_{\rm SDL}$ respectively.

	We compare these approximations with a numerical solution to the exact lens equation (\ref{lenseq_exact}), which is denoted by $x_{\rm num}$,  
	and discuss the accuracy of the image position $\Theta$.
	As an example, we take $u_{\rm O}=2M/r_{\rm O}=5\times10^{-11}$, which roughly corresponds to the supermassive black hole at the center of the Galaxy, i.e. $M\sim4\times10^6M_{\odot}, D_{\rm OL}\sim8\rm kpc$.
	For simplicity, we also assume $u_{\rm S}=5\times10^{-11}$.
	We plotted the approximate solutions (\ref{xO(1)}), (\ref{xSDL}) and their relative errors $|1-x_{O(1)}/x_{\rm num}|$,
	$|1-x_{\rm SDL}/x_{\rm num}|$ for $5\pi/4\leq\phi\leq7\pi$ in Figure \ref{fig:lenssol}.
	Note that for small $u_{\rm O}$ the relative error of the image position $\delta\Theta$ is related to that of $x$ as
	$|\delta\Theta/\Theta|\sim|\delta x/2x|$.
	For the Schwarzschild black hole case, the relative error to the Einstein ring position of the winding number $N=1$ ($\phi=3\pi$ in our notation)
	using the standard logarithmic formula was shown to be about $10^{-4}$\cite{Bozza_2001}.
	In order to achieve the same level of accuracy at $\phi=3\pi$, 
	Figure \ref{fig:lenssol} shows that the approximation $x_{O(1)}$ is not sufficient but $x_{\rm SDL}$ is necessary.

	\begin{figure}[htb]
		\centering
		\begin{minipage}{0.49\linewidth}
			\centering
			\includegraphics[keepaspectratio,width=0.9\linewidth]{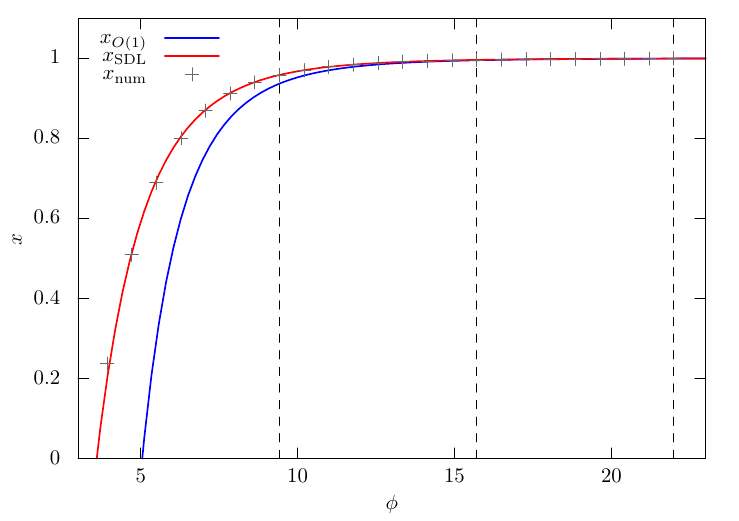}
			\label{fig:x_phi}
		\end{minipage}
		\begin{minipage}{0.49\linewidth}
			\centering
			\includegraphics[keepaspectratio,width=0.9\linewidth]{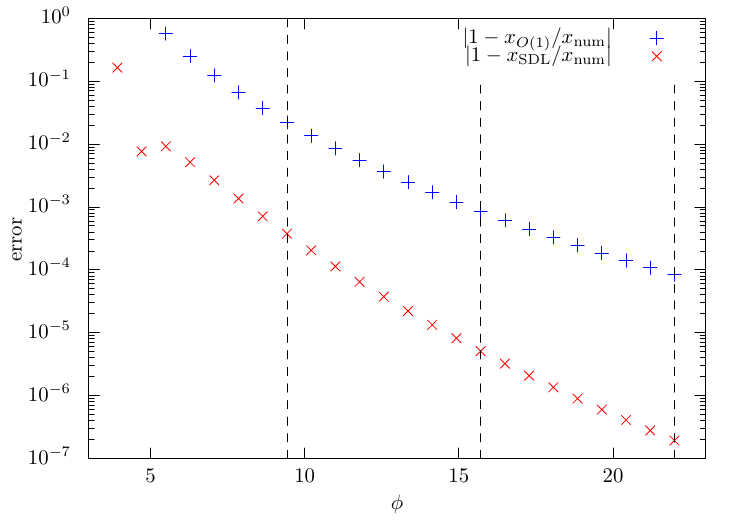}
			\label{fig:x_phi_RE}
		\end{minipage}
		\caption{
			The approximate solutions (\ref{xO(1)}) and (\ref{xSDL}) to the lens equation (left panel) 
			and the abosolute value of the relative errors (right panel) compared with the numerical solution
			in the SDL from outside the photon sphere are plotted. 
			The vertical dashed lines represent the positions of the Einstein rings $\phi=(2N+1)\pi$ with the winding number $N=1,2,3$.
		}
		\label{fig:lenssol}
	\end{figure}

	We next consider the SDL from inside the photon sphere $(x\to1+0)$.
	In this case, we can approximate the lens equation as
	\begin{equation}
		\phi=\sqrt{3}a_0\left(1-\frac{1}{x}\right)^{-1/6}-c_0-\sqrt{3}b_0\left(1-\frac{1}{x}\right)^{1/6}.
	\end{equation}
	The solution is given by
	\begin{equation}
		\frac{1}{x}=1-\left(\frac{2\sqrt{3}a_0}{\phi+c_0}\right)^6\left(1+\sqrt{1+\frac{12a_0b_0}{(\phi+c_0)^2}}\right)^{-6},\label{xinvSDL}
	\end{equation}
	which is again denoted by $1/x_{\rm SDL}$.
	This solution and the numerical solution are plotted in Figure \ref{fig:lenssol_inside}.
	In contrast to the SDL from outside, the approximate solution (\ref{xinvSDL}) has an error of more than $10\%$ 
	for the image with the winding number $N=1$ ($\Leftrightarrow\phi=3\pi$) while the error becomes less than $1\%$ for the image with $N\geq2$.
	For more accuracy, one has to include higher-order terms in the lens equation to approximate the deflection angle.
	Nevertheless, we summarize the leading three ring positions in Table \ref{table_EinsteinRing}.
	
	\begin{table}[htbp]
		\centering
		\caption{
			The diameters of the relativistic Einstein rings for the winding numbers $N=1,2,3$ calculated by using the SDL formulae
			(\ref{xSDL}) and (\ref{xinvSDL}) compared with the numerical solutions. The superscripts (out) and (in) represent
			outside and inside the photon sphere respectively.
		}
		\label{table_EinsteinRing}
		\begin{tabular}{c|ccc}
			& $N=1$ & $N=2$ & $N=3$ \\\hline\hline
			$2\Theta_{\rm SDL}^{\rm (out)}/\mu {\rm as}$ & $38.7085$ & $37.9711$ & $37.9074$ \\
			$2\Theta_{\rm num}^{\rm (out)}/\mu {\rm as}$ & $38.7158$ & $37.9712$ & $37.9074$ \\ \hline
			$2\Theta_{\rm SDL}^{\rm (in)}/\mu {\rm as}$ & $24.2763$ & $36.3597$ & $37.5779$ \\
			$2\Theta_{\rm num}^{\rm (in)}/\mu {\rm as}$ & $30.2044$ & $36.4955$ & $37.5845$ 
		\end{tabular}
	\end{table}

	\begin{figure}[htb]
		\centering
		\begin{minipage}{0.49\linewidth}
			\centering
			\includegraphics[keepaspectratio,width=0.9\linewidth]{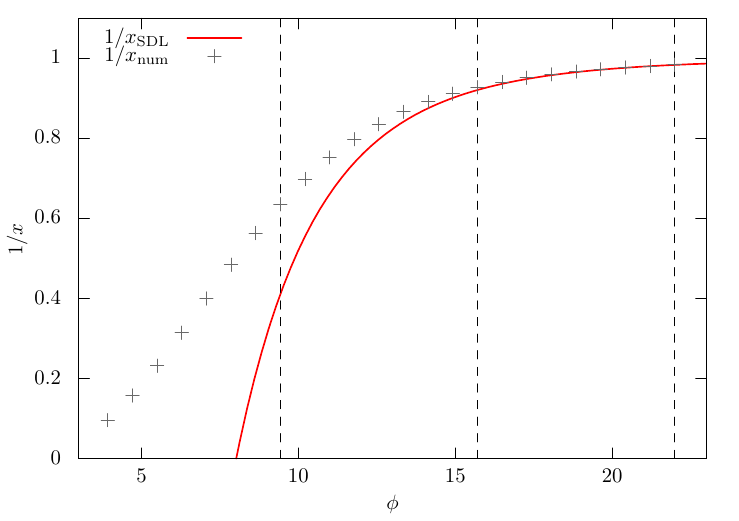}
			\label{fig:x_phi_inside}
		\end{minipage}
		\begin{minipage}{0.49\linewidth}
			\centering
			\includegraphics[keepaspectratio,width=0.9\linewidth]{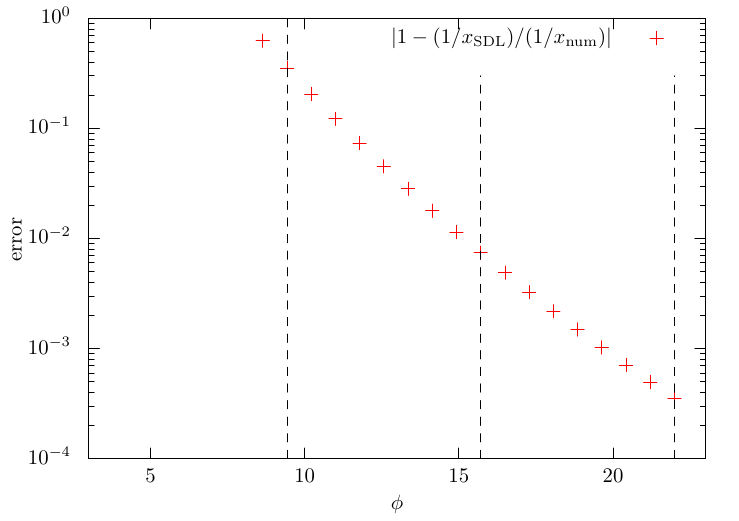}
			\label{fig:x_phi_RE_inside}
		\end{minipage}
		\caption{
			The approximate solution (\ref{xinvSDL}) to the lens equation (left panel) 
			and the abosolute value of the relative error (right panel) compared with the numerical solution 
			in the SDL from inside the photon sphere are plotted. 
			The vertical dashed lines represent the positions of the Einstein rings $\phi=(2N+1)\pi$ with the winding number $N=1,2,3$.
		}
		\label{fig:lenssol_inside}
	\end{figure}

}}}

\section{Discussion\label{sec_discussion}}
{{{
	In this paper, we have solved the PF equation (\ref{PFeq}) satisfied by the deflection angle $\Delta\phi(x)$ as a function of the impact parameter $x=(b_{\rm m}/b)^2$
	and have obtained the full-order power series expressions (\ref{dphiWDL}), (\ref{dphiSDLoutside}), (\ref{dphiSDLinside}), and (\ref{dphiinfinity}),
	which are valid in the WDL $(x\to+0)$, SDL from the outside $(x\to1-0)$, SDL from the inside $(x\to1+0)$, and in the head-on limit $(x\to+\infty)$, respectively.
	Note that (\ref{dphiWDL}) is a specific case for $Q^2=9M^2$ of the generic expression derived in \cite{Sasaki_2024}.
	By truncating these exact expressions at a finite order, we can construct an approximate expression for $\Delta\phi(x)$ 
	with desired accuracy as shown in Figures \ref{fig_dphi} and \ref{fig_error}.
	In particular, the asymptotic formula derived from (\ref{dphiSDLoutside}) in the SDL from outside the photon sphere 
	by truncating up to the $O(1)$-term is compared with the result by Tsukamoto\cite{Tsukamoto2020}.
	Observing the asymptotic behavior of the relative errors with the numerical solution in the limit $x\to1-0\ (\Leftrightarrow b\to b_{\rm m}+0)$,
	we concluded that the numerical coefficient $\bar{c}$ of the divergent term given by (\ref{cf_Tsukamoto2020})\cite{Tsukamoto2020} is incorrect
	though the $O(1)$-term is consistent with our result.
	The asymptotic formula (\ref{n=0SDLinside}) derived by truncating the expansion (\ref{dphiSDLinside}) in the SDL from the inside 
	appears in the literature for the first time.

	The logarithmic formula in the Reissner-Nordstr\"{o}m naked singularity for $1<Q^2/M^2<9/8$ has been derived by Tsukamoto\cite{Tsukamoto_2017,Tsukamoto_2021}.
	However, the formulae given in these works break down in the marginal limit $Q^2/M^2\to9/8$ as explicitly mentioned in \cite{Tsukamoto_2021}.
	Our formulae (\ref{dphiSDLoutside}) and (\ref{dphiSDLinside}) partially fill a gap in this situation
	but neither of the formulae in \cite{Tsukamoto_2021} nor ours can give an asymptotic expression 
	when $Q^2/M^2$ is very close to but not equal to $9/8$.
	Note that this problem cannot be resolved by adding higher-order terms to the logarithmic formula as the coefficients
	of such terms diverge strongly in this limit\cite{Sasaki_2024}.

	In the literature\cite{Paul2020,Tsukamoto2020b,Tsukamoto2020,Tsukamoto2021}, the deflection angles that are non-logarithmically divergent in 
	the strong deflection limit have been approximated by the leading divergent term and the next-leading $O(1)$-term.
	However, we have shown that such a formula (at least for our case) is less accurate than the standard formula\cite{Bozza_2002,Tsukamoto_2017}
	used when the deflection angle diverges logarithmically
	so that one has to include higher-order terms to approximate the deflection angle to achieve the same level of accuracy.
	For the SDL from the outside, it is sufficient to include the next order term, i.e. $O((b/b_{\rm m}-1)^{1/6})$-term
	while the approximation with the same order for the SDL from the inside has a larger error.

	The power series expression (\ref{dphiinfinity}) for the deflection angle in the head-on limit $x\to+0\ (\Leftrightarrow b/b_{\rm m}\to+0)$ 
	can be used to investigate the scattering of massless particles that pass very close to the naked singularity.
	In this limit, the deflection angle $\Delta\phi$ converges to $0$ as $\Delta\phi\sim x^{-1/4}\propto b^{1/2}$
	so that the particle is scattered back to the source.

	Lastly, we solved the lens equation (\ref{Lenseq_far}) by using the $(n=0)$-formula for the deflection angle $\Delta\phi$
	in the SDL both from outside and inside the photon sphere.
	The image position is given by $\Theta\simeq\sqrt{27/8x}u_{\rm O}$ with (\ref{xSDL}) or (\ref{xinvSDL}).
	Although these solutions are derived under the assumption that both the source and the observer lie far away from the lens,
	the configuration is not restricted to the almost perfectly aligned case, i.e. $\phi\sim(2N+1)\pi$.
	For example, the leading contribution to the retro-lensing corresponds to the images around $\phi=2\pi$.
	As shown in Figure \ref{fig:lenssol}, the approximation (\ref{xSDL}) is valid around $\phi=2\pi$ with the error of $10^{-2}$.
	On the other hand, the approximation (\ref{xinvSDL}) cannot be used there due to a large error as shown in Figure \ref{fig:lenssol_inside}.

}}}

\paragraph{Data Availability Statement}
The data that support the findings of this study are openly available on the author's researchmap page\cite{Dataset_Sasaki2025}.

\appendix
\section{Analytic continuation of $\Delta\phi_{\rm I}^{(0)}$ and $\Delta\phi_{\rm I}^{(\infty)}(x)$ to $x=1$}
	\subsection{Derivation of eq.(\ref{dphi0_continuation})\label{Derivation1}}
	{{{
		We start from eq.(\ref{dphi0C}).
		Since $\omega_1(x)$ and $\omega_2(x)$ can be expanded in terms of $(1-x)^{n-1/6}$ and $(1-x)^{n+1/6}$ respectively,
		we evaluate the following integrals:
		\begin{equation}
			J_1^{(n)}(x):=\int_0^x\frac{\mathrm{d}x'}{\sqrt{x'}}\left(\frac{4}{3}-x'\right)(1-x')^{n-7/6},\ \ 
			J_2^{(n)}(x):=\int_0^x\frac{\mathrm{d}x'}{\sqrt{x'}}\left(\frac{4}{3}-x'\right)(1-x')^{n-5/6},
		\end{equation}
		for $n=0,1,2,\cdots$ in the limit $x\to1-0$.
		In order to explicitly derive the coefficients for $\Delta\phi_{\rm I}^{(0)}$ up to $O((1-x)^{1/6})$,
		we have to evaluate $J_1^{(n)}(x)$ and $J_2^{(n)}(x)$ up to $O(1)$ and $O((1-x)^{1/6})$ respectively.
		Among $J_1^{(n)}(x)$, only $J_1^{(0)}(x)$ is divergent in $x\to1-0$ so we perform the integration before taking the limit,
		\begin{align}
			J_1^{(0)}(x)=&\frac{4}{3}x^{1/2}\int_0^1\frac{\mathrm{d}t}{t^{1/2}}(1-xt)^{-7/6}-x^{3/2}\int_0^1t^{1/2}(1-xt)^{-7/6}\mathrm{d}t\ \ (x'=xt)\notag\\
			=&\frac{8}{3}x^{1/2}{}_2F_1\left[\begin{matrix}\frac{7}{6},\frac{1}{2}\\ \frac{3}{2}\end{matrix};x\right]
				-\frac{2}{3}x^{3/2}{}_2F_1\left[\begin{matrix}\frac{7}{6},\frac{3}{2}\\ \frac{5}{2}\end{matrix};x\right].
		\end{align}
		By using the continuation formula\cite{HTF1},
		\begin{align}
			{}_2F_1\left[\begin{matrix}a,b\\ c\end{matrix};x\right]=&\frac{\Gamma(c)\Gamma(c-a-b)}{\Gamma(c-a)\Gamma(c-b)}{}_2F_1\left[\begin{matrix}a,b\\ a+b-c+1\end{matrix};1-x\right]\notag\\
			&+\frac{\Gamma(c)\Gamma(a+b-c)}{\Gamma(a)\Gamma(b)}\left(1-x\right)^{c-a-b}{}_2F_1\left[\begin{matrix}c-a,c-b\\c-a-b+1\end{matrix};1-x\right],
		\end{align}
		we obtain
		\begin{equation}
			J_1^{(0)}(x)=2(1-x)^{-1/6}+\frac{2\pi^{3/2}}{\Gamma(\frac{1}{3})\Gamma(\frac{1}{6})}+O((1-x)^{5/6}),\ \ x\to1-0.
		\end{equation}
		Since $J_1^{(n)}(x)$ are finite in the limit $x\to1-0$ for $n\geq1$, we only evaluate $J_1^{(n)}(1)$ as
		\begin{align}
			J_1^{(n)}(1)=&\frac{4}{3}\int_0^1\left(1-\frac{3x}{4}\right)x^{-1/2}(1-x)^{n-7/6}\mathrm{d}x\notag\\
			=&\frac{(24n-1)\Gamma(\frac{1}{2})\Gamma(n-\frac{1}{6})}{18\Gamma(n+\frac{4}{3})}.
		\end{align}
		Note that $O(1)$-term in $J_1^{(0)}(x)$ is equal to the limti $\lim_{n\to0}J_1^{(n)}(1)$.
		Similarly, $J_2^{(n)}(1)\ (n\geq0)$ can be evaluated as
		\begin{align}
			J_2^{(n)}(1)=&\frac{4}{3}\int_0^1\left(1-\frac{3x}{4}\right)x^{-1/2}(1-x)^{n-5/6}\mathrm{d}x\\
			=&\frac{(24n+7)\Gamma(\frac{1}{2})\Gamma(n+\frac{1}{6})}{18\Gamma(n+\frac{5}{3})}.
		\end{align}
		Among $J_2^{(n)}(x)$, only $J_2^{(0)}(x)$ contributes to $O((1-x)^{1/6})$.
		To determine this contribution, we expand the integrand around $x'=1$ and integrate the leading order term,
		\begin{align}
			J_2^{(0)}(x)=&\int_0^x\mathrm{d}x'\frac{1}{3}(1-x')^{-5/6}\left(1+O(1-x')\right)\notag\\
			=&-2(1-x)^{1/6}+O(1).
		\end{align}
		Using the results above, we manipulate (\ref{dphi0C}) as
		\begin{align}
			\Delta\phi_{\rm I}^{(0)}(x)=&\frac{\sqrt{6}}{8}\left[\omega_1(x)\sum_{n=0}^\infty\frac{(\frac{5}{12})_n(\frac{11}{12})_n}{n!(\frac{4}{3})_n}J_2^{(n)}(x)
				-\omega_2(x)\sum_{n=0}^\infty\frac{(\frac{1}{12})_n(\frac{7}{12})_n}{n!(\frac{2}{3})_n}J_1^{(n)}(x)\right]\notag\\
			=&\frac{\sqrt{6}}{8}\Biggl[(1-x)^{-1/6}(1+O(1-x))\Biggl\{-2(1-x)^{1/6}\notag\\
				&\hspace{4em}+\sum_{n=0}^\infty\frac{(\frac{5}{12})_n(\frac{11}{12})_n}{n!(\frac{4}{3})_n}\frac{(24n+7)\Gamma(\frac{1}{2})\Gamma(n+\frac{1}{6})}{18\Gamma(n+\frac{5}{3})}+O((1-x)^{7/6})\Biggr\}\notag\\
				&\hspace{2em}-(1-x)^{1/6}(1+O(1-x))\Biggl\{2(1-x)^{-1/6}\notag\\
				&\hspace{4em}+\sum_{n=0}^\infty\frac{(\frac{1}{12})_n(\frac{7}{12})_n}{n!(\frac{2}{3})_n}\frac{(24n-1)\Gamma(\frac{1}{2})\Gamma(n-\frac{1}{6})}{18\Gamma(n+\frac{4}{3})}+O((1-x)^{5/6})\Biggr\}\Biggr]\notag\\
			=&N_1\cdot(1-x)^{-1/6}(1+O(1-x))-N_2\cdot(1-x)^{1/6}(1+O(1-x))-\frac{\sqrt{6}}{2}(1+O(1-x)),
		\end{align}
		where $N_{1,2}$ are given by (\ref{N1}) and (\ref{N2}).
	}}}
	\subsection{Derivation of eq.(\ref{dphiinf_continuation})\label{Derivation2}}
	{{{
		We start from (\ref{dphiinfC}) and (\ref{Ceqinf}).
		We first derive $1/x$ expansions of $\omega_i^{(\infty)}(x)=(x-1)^{-1/6}f_i^{(\infty)}(x)\ (i=1,2)$ as follows:
		\begin{align}
			\omega_1^{(\infty)}(x)=&x^{-1/4}\left(1-\frac{1}{x}\right)^{-1/6}{}_2F_1\left[\begin{matrix}\frac{1}{12},\frac{1}{12}\\\frac{1}{2}\end{matrix};\frac{1}{x}\right]\notag\\
				=&x^{-1/4}\sum_{n=0}^\infty\frac{(\frac{1}{6})_n}{n!}x^{-n}\sum_{j=0}^\infty\frac{(\frac{1}{12})_j(\frac{1}{12})_j}{j!(\frac{1}{2})_j}x^{-j}\notag\\
				=&x^{-1/4}\sum_{p=0}^\infty x^{-p}\sum_{j=0}^p\frac{(\frac{1}{12})_j(\frac{1}{12})_j}{j!(\frac{1}{2})_j}\frac{(\frac{1}{6})_{p-j}}{(p-j)!}\notag\\
				=&x^{-1/4}\sum_{p=0}^\infty \frac{(\frac{1}{6})_p}{p!}x^{-p}{}_3F_2\left[\begin{matrix}-p,\frac{1}{12},\frac{1}{12}\\-p+\frac{5}{6},\frac{1}{2}\end{matrix};1\right],
		\end{align}
		and similarly
		\begin{align}
			\omega_2^{(\infty)}(x)=&x^{-3/4}\left(1-\frac{1}{x}\right)^{-1/6}{}_2F_1\left[\begin{matrix}\frac{7}{12},\frac{7}{12}\\\frac{3}{2}\end{matrix};\frac{1}{x}\right]\notag\\
				=&x^{-3/4}\sum_{p=0}^\infty\frac{(\frac{1}{6})_p}{p!}x^{-p}{}_3F_2\left[\begin{matrix}-p,\frac{7}{12},\frac{7}{12}\\-p+\frac{5}{6},\frac{3}{2}\end{matrix};1\right].
		\end{align}
		In these derivation, we have changed the variable of the sum as $n\to p=n+j$.
		Using these expressions, the term by term integrations of (\ref{Ceqinf}) give
		\begin{align}
			C_1(x)=&-\frac{\sqrt{6}}{36}\int\frac{\mathrm{d}x}{\sqrt{x}}\frac{3x-4}{x-1}\omega_2^{(\infty)}(x)\notag\\
				=&-\frac{\sqrt{6}}{12}\sum_{p=0}^\infty\frac{(\frac{1}{6})_p}{p!}
				{}_3F_2\left[\begin{matrix}-p,\frac{7}{12},\frac{7}{12}\\-p+\frac{5}{6},\frac{3}{2}\end{matrix};1\right]
					\int\frac{\mathrm{d}x}{\sqrt{x}}\frac{1-4/3x}{1-1/x}x^{-p-3/4}\notag\\
				=&-\frac{\sqrt{6}}{12}\sum_{p=0}^\infty\frac{(\frac{1}{6})_p}{p!}
				{}_3F_2\left[\begin{matrix}-p,\frac{7}{12},\frac{7}{12}\\-p+\frac{5}{6},\frac{3}{2}\end{matrix};1\right]
					\sum_{j=0}^\infty\int\mathrm{d}x\left(1-\frac{4}{3x}\right)x^{-p-j-5/4}\notag\\
				=&-\frac{\sqrt{6}}{12}\sum_{p=0}^\infty\frac{(\frac{1}{6})_p}{p!}
				{}_3F_2\left[\begin{matrix}-p,\frac{7}{12},\frac{7}{12}\\-p+\frac{5}{6},\frac{3}{2}\end{matrix};1\right]
					\sum_{j=0}^\infty\left(-\frac{x^{-p-j-1/4}}{p+j+1/4}+\frac{4}{3}\frac{x^{-p-j-5/4}}{p+j+5/4}\right)\notag\\
				=&\frac{\sqrt{6}}{12}\Biggl(4x^{-1/4}+x^{-5/4}\sum_{n=0}^\infty\frac{x^{-n}}{n+5/4}\left(\sum_{p=0}^{n+1}-\frac{4}{3}\sum_{p=0}^{n}\right)\frac{(\frac{1}{6})_p}{p!}{}_3F_2\left[\begin{matrix}-p,\frac{7}{12},\frac{7}{12}\\-p+\frac{5}{6},\frac{3}{2}\end{matrix};1\right]\Biggr),
		\end{align}
		and similarly
		\begin{align}
			C_2(x)=&\frac{\sqrt{6}}{36}\int\frac{\mathrm{d}x}{\sqrt{x}}\frac{3x-4}{x-1}\omega_1^{(\infty)}(x)\notag\\
				=&\frac{\sqrt{6}}{12}\sum_{p=0}^\infty\frac{(\frac{1}{6})_p}{p!}
					{}_3F_2\left[\begin{matrix}-p,\frac{1}{12},\frac{1}{12}\\-p+\frac{5}{6},\frac{1}{2}\end{matrix};1\right]
					\int\frac{\mathrm{d}x}{\sqrt{x}}\frac{1-4/3x}{1-1/x}x^{-p-1/4}\notag\\
				=&\frac{\sqrt{6}}{12}\Biggl(4x^{1/4}-x^{-3/4}\sum_{n=0}^\infty\frac{x^{-n}}{n+3/4}\left(\sum_{p=0}^{n+1}-\frac{4}{3}\sum_{p=0}^n\right)\frac{(\frac{1}{6})_p}{p!}{}_3F_2\left[\begin{matrix}-p,\frac{1}{12},\frac{1}{12}\\-p+\frac{5}{6},\frac{1}{2}\end{matrix};1\right]\Biggr).
		\end{align}
		Using these expressions, we can confirm the boundary condition at $x=\infty$ as
		\begin{equation}
			C_1(x)\omega_1^{(\infty)}(x)+C_2(x)\omega_2^{(\infty)}(x)=\frac{2\sqrt{6}}{3\sqrt{x}}+\cdots=\Delta\phi_{\rm I}^{(\infty)}(x).
		\end{equation}

		We next perform the integration for $C_i(x)$ using the continuation formulae (\ref{infto1_1}) and (\ref{infto1_2}).
		\begin{align}
			C_1(x)=&-\frac{\sqrt{6}}{36}\int\frac{\mathrm{d}x}{\sqrt{x}}\frac{3x-4}{(x-1)^{7/6}}\left(\frac{\pi^{3/2}}{\sqrt{3}\Gamma^2(\frac{11}{12})\Gamma(\frac{2}{3})}\tilde{f}_1^{(1)}(x)
				-\frac{\pi^{3/2}}{\sqrt{3}\Gamma^2(\frac{7}{12})\Gamma(\frac{4}{3})}\tilde{f}_2^{(1)}(x)\right),\\
			C_2(x)=&\frac{\sqrt{6}}{36}\int\frac{\mathrm{d}x}{\sqrt{x}}\frac{3x-4}{(x-1)^{7/6}}\left(\frac{2\pi^{3/2}}{\sqrt{3}\Gamma^2(\frac{5}{12})\Gamma(\frac{2}{3})}\tilde{f}_1^{(1)}(x)
				-\frac{2\pi^{3/2}}{\sqrt{3}\Gamma^2(\frac{1}{12})\Gamma(\frac{4}{3})}\tilde{f}_2^{(1)}(x)\right).
		\end{align}
		We again perform term by term integration by expanding $\tilde{f}_i^{(1)}(x)$ in terms of $1-1/x$.
		For $\tilde{f}_1^{(1)}(x)$, we proceed as follows:
		\begin{align}
			\int\frac{\mathrm{d}x}{\sqrt{x}}\frac{3x-4}{(x-1)^{7/6}}\tilde{f}_1^{(1)}(x)
			=&\sum_{n=0}^\infty\frac{(\frac{1}{12})_n(\frac{1}{12})_n}{n!(\frac{2}{3})_n}\int\mathrm{d}x\left(3-\frac{4}{x}\right)x^{-3/4}\left(1-\frac{1}{x}\right)^{n-7/6}\notag\\
			=&\sum_{n=0}^\infty\frac{(\frac{1}{12})_n(\frac{1}{12})_n}{n!(\frac{2}{3})_n}\left(12x^{1/4}{}_2F_1\left[\begin{matrix}-n+\frac{7}{6},-\frac{1}{4}\\\frac{3}{4}\end{matrix};\frac{1}{x}\right]
				+\frac{16}{3}x^{-3/4}{}_2F_1\left[\begin{matrix}-n+\frac{7}{6},\frac{3}{4}\\\frac{7}{4}\end{matrix};\frac{1}{x}\right]\right).
		\end{align}
		Similarly, we obtain for $\tilde{f}_2^{(1)}(x)$
		\begin{align}
			\int\frac{\mathrm{d}x}{\sqrt{x}}\frac{3x-4}{(x-1)^{7/6}}\tilde{f}_2^{(1)}(x)
			=&\sum_{n=0}^\infty\frac{(\frac{5}{12})_n(\frac{5}{12})_n}{n!(\frac{4}{3})_n}\int\mathrm{d}x\left(3-\frac{4}{x}\right)x^{-3/4}\left(1-\frac{1}{x}\right)^{n-5/6}\notag\\
			=&\sum_{n=0}^\infty\frac{(\frac{5}{12})_n(\frac{5}{12})_n}{n!(\frac{4}{3})_n}\left(12x^{1/4}{}_2F_1\left[\begin{matrix}-n+\frac{5}{6},-\frac{1}{4}\\\frac{3}{4}\end{matrix};\frac{1}{x}\right]
				+\frac{16}{3}x^{-3/4}{}_2F_1\left[\begin{matrix}-n+\frac{5}{6},\frac{3}{4}\\\frac{7}{4}\end{matrix};\frac{1}{x}\right]\right).
		\end{align}
		In these infinite series, we use the following continuation formula for ${}_2F_1$\cite{HTF1}:
		\begin{align}
			{}_2F_1\left[\begin{matrix}a,b\\ c\end{matrix};\frac{1}{x}\right]=&\frac{\Gamma(c)\Gamma(c-a-b)}{\Gamma(c-a)\Gamma(c-b)}{}_2F_1\left[\begin{matrix}a,b\\ a+b-c+1\end{matrix};1-\frac{1}{x}\right]\notag\\
			&+\frac{\Gamma(c)\Gamma(a+b-c)}{\Gamma(a)\Gamma(b)}\left(1-\frac{1}{x}\right)^{c-a-b}{}_2F_1\left[\begin{matrix}c-a,c-b\\c-a-b+1\end{matrix};1-\frac{1}{x}\right].
		\end{align}
		Although this formula enables us to derive full order expressions for $C_i(x)\ (i=1,2)$ around $x=1$, we here only show the first several terms necessary to determine the connection coefficients.
		\begin{align}
			\int\frac{\mathrm{d}x}{\sqrt{x}}\frac{3x-4}{(x-1)^{7/6}}\tilde{f}_1^{(1)}(x)=&\tilde{N}_1+6(x-1)^{-1/6}+O((x-1)^{5/6}),\\
			\int\frac{\mathrm{d}x}{\sqrt{x}}\frac{3x-4}{(x-1)^{7/6}}\tilde{f}_2^{(1)}(x)=&\tilde{N}_2-6(x-1)^{1/6}+O((x-1)^{7/6}),
		\end{align}
		where constants $\tilde{N}_i\ (i=1,2)$ are given by
		\begin{align}
			\tilde{N}_1=&\frac{6\Gamma(\frac{3}{4})\Gamma(\frac{5}{6})}{\Gamma(\frac{7}{12})}\left(5\cdot{}_3F_2\left[\begin{matrix}\frac{1}{12},\frac{1}{12},-\frac{1}{6}\\\frac{2}{3},-\frac{5}{12}\end{matrix};1\right]
				-4\cdot{}_3F_2\left[\begin{matrix}\frac{1}{12},\frac{1}{12},-\frac{1}{6}\\\frac{2}{3},\frac{7}{12}\end{matrix};1\right]\right),\label{Nt1}\\
			\tilde{N}_2=&\frac{\Gamma(\frac{3}{4})\Gamma(\frac{1}{6})}{\Gamma(\frac{11}{12})}\left(4\cdot{}_3F_2\left[\begin{matrix}\frac{5}{12},\frac{5}{12},\frac{1}{6}\\\frac{4}{3},\frac{11}{12}\end{matrix};1\right]
				-{}_3F_2\left[\begin{matrix}\frac{5}{12},\frac{5}{12},\frac{1}{6}\\\frac{4}{3},-\frac{1}{12}\end{matrix};1\right]\right).\label{Nt2}
		\end{align}
		Then, the asymptotic forms of $C_i\ (i=1,2)$ become
		\begin{align}
			C_1(x)=&-\frac{\sqrt{2}}{6}\frac{\pi^{3/2}}{\Gamma^2(\frac{11}{12})\Gamma(\frac{2}{3})}(x-1)^{-1/6}-\frac{\sqrt{2}}{6}\frac{\pi^{3/2}}{\Gamma^2(\frac{7}{12})\Gamma(\frac{4}{3})}(x-1)^{1/6}\notag\\
			&-\frac{\sqrt{2}\pi^{3/2}}{36}\left(\frac{\tilde{N}_1}{\Gamma^2(\frac{11}{12})\Gamma(\frac{2}{3})}-\frac{\tilde{N}_2}{\Gamma^2(\frac{7}{12})\Gamma(\frac{4}{3})}\right)+O((x-1)^{5/6}),\\
			C_2(x)=&\frac{\sqrt{2}}{3}\frac{\pi^{3/2}}{\Gamma^2(\frac{5}{12})\Gamma(\frac{2}{3})}(x-1)^{-1/6}+\frac{\sqrt{2}}{3}\frac{\pi^{3/2}}{\Gamma^2(\frac{1}{12})\Gamma(\frac{4}{3})}(x-1)^{1/6}\notag\\
			&+\frac{\sqrt{2}\pi^{3/2}}{18}\left(\frac{\tilde{N}_1}{\Gamma^2(\frac{5}{12})\Gamma(\frac{2}{3})}-\frac{\tilde{N}_2}{\Gamma^2(\frac{1}{12})\Gamma(\frac{4}{3})}\right)+O((x-1)^{5/6}).
		\end{align}
		Inserting these expressions into the inhomogeneous solution (\ref{dphiinfC}), we finally obtain 
		\begin{equation}
			\Delta\phi_{\rm I}^{(\infty)}(x)=\frac{\sqrt{6}}{24}\tilde{N}_2(x-1)^{-1/6}(1+O(x-1))-\frac{\sqrt{6}}{24}\tilde{N}_1(x-1)^{1/6}(1+O(x-1))-\frac{\sqrt{6}}{2}(1+O(x-1)),\ (x\to1+0)
		\end{equation}
	}}}

\section{Explicit coefficients at each singularity\label{appendix_coeff}}
{{{
	In this appendix, we show explicit series expressions for $\Delta\phi$ at singularities $x=0,1$, and $\infty$
	up to the order necessary to recover the results in figures \ref{fig_dphi} and \ref{fig_error}.
	Since $\Delta\phi$ is discontinuous at $x=1$, we separately show the right $x\to1+0$ and the left $x\to1-0$ limits.
	\subsection{$x=0$ (WDL)}
		The exact expression (\ref{dphiWDL}) can be expanded as
		\begin{align}
			\Delta\phi(x)=&\frac{\pi}{2}\Bigl[1+\frac{31}{\num{144}}x+\frac{\num{10465}}{\num{82944}}x^2+\frac{\num{9769375}}{\num{107495424}}x^3+\frac{\num{17761128385}}{\num{247669456896}}x^4+\frac{\num{2124864853111}}{\num{35664401793024}}x^5\notag\\
			&+\frac{\num{9459850016097481}}{\num{184884258895036416}}x^6
			+\frac{\num{1197199678487902375}}{\num{26623333280885243904}}x^7
			+\frac{\num{9863295534688132504825}}{\num{245360639516638407819264}}x^8+O(x^9)
			\Bigr]\notag\\
			&+\sqrt{\frac{8x}{27}}\Bigl[1+\frac{37}{81}x+\frac{\num{1109}}{\num{3645}}x^2+\frac{\num{53063}}{\num{229635}}x^3+\frac{\num{6283807}}{\num{33480783}}x^4\notag
			+\frac{\num{75189623}}{\num{473513931}}x^5
			+\frac{\num{7649949023}}{\num{55401129927}}x^6\notag\\
			&+\frac{\num{2747832171559}}{\num{22437457620435}}x^7
			+\frac{\num{2646493260181}}{\num{24006510600885}}x^8+O(x^9)\Bigr].
		\end{align}
		
	\subsection{$x\to1-0$ (SDL from outside)}
		The exact expression (\ref{dphiSDLoutside}) can be expanded as
		\begin{align}
			\Delta\phi(x)=&\frac{\Gamma(\frac{1}{12})\Gamma(\frac{7}{12})}{2\sqrt{3}\Gamma(\frac{2}{3})}(1-x)^{-1/6}\Bigl[1+\frac{7(1-x)}{96}
			+\frac{\num{1729}(1-x)^2}{\num{46080}}+\frac{\num{267995}(1-x)^3}{\num{10616832}}+\frac{\num{426380045}(1-x)^4}{\num{22422749184}}\notag\\
			&+\frac{\num{2984660315}(1-x)^5}{\num{195689447424}}+\frac{\num{12198306707405}(1-x)^6}{\num{958095534587904}}
			+\frac{\num{2009932422331561}(1-x)^7}{\num{183954342640877568}}\notag\\
			&+\frac{\num{914519252160860255}(1-x)^8}{\num{95569691423778275328}}+O((1-x)^9)\Bigr]\notag\\
			&-\frac{\Gamma(\frac{5}{12})\Gamma(\frac{11}{12})}{2\sqrt{3}\Gamma(\frac{4}{3})}(1-x)^{1/6}\Bigl[1+\frac{55(1-x)}{192}
			+\frac{\num{21505}(1-x)^2}{\num{129024}}+\frac{\num{623645}(1-x)^3}{\num{5308416}}+\frac{\num{1201763915}(1-x)^4}{\num{13249806336}}\notag\\
			&+\frac{\num{751583152441}(1-x)^5}{\num{10175851266048}}+\frac{\num{266812019116555}(1-x)^6}{\num{4283250625216512}}
			+\frac{\num{22145397586674065}(1-x)^7}{\num{411192060020785152}}\notag\\
			&+\frac{\num{394188077042798357}(1-x)^8}{\num{8310407949893763072}}+O((1-x)^9)\Bigr]\notag\\
			&-\frac{\sqrt{6}}{2}\Bigl[1+\frac{13(1-x)}{140}+\frac{87(1-x)^2}{\num{2288}}+\frac{\num{5851}(1-x)^3}{\num{268736}}+\frac{\num{137951}(1-x)^4}{\num{9509120}}
			+\frac{\num{1115979}(1-x)^5}{\num{105866240}}\notag\\
			&+\frac{\num{27596253}(1-x)^6}{\num{3406131200}}+\frac{\num{6222910719}(1-x)^7}{\num{960801488896}}
			+\frac{\num{52480199997}(1-x)^8}{\num{9845276213248}}+O((1-x)^9)\Bigr].
		\end{align}

	\subsection{$x\to1+0$ (SDL from inside)}
		The exact expression (\ref{dphiSDLinside}) is expanded in terms of $1-1/x$ as
		\begin{align}
			\Delta\phi(x)=&\frac{\Gamma(\frac{1}{12})\Gamma(\frac{7}{12})}{2\Gamma(\frac{2}{3})}\left(1-\frac{1}{x}\right)^{-1/6}
			\Bigl[1-\frac{23}{96}\left(1-\frac{1}{x}\right)-\frac{\num{4271}}{\num{46080}}\left(1-\frac{1}{x}\right)^2
			-\frac{\num{2897927}}{\num{53084160}}\left(1-\frac{1}{x}\right)^3\notag\\
			&-\frac{\num{4228798559}}{\num{112113745920}}\left(1-\frac{1}{x}\right)^4
			-\frac{\num{305427860509}}{\num{10762919608320}}\left(1-\frac{1}{x}\right)^5
			-\frac{\num{1186429157747749}}{\num{52695254402334720}}\left(1-\frac{1}{x}\right)^6\notag\\
			&-\frac{\num{187422445652646191}}{\num{10117488845248266240}}\left(1-\frac{1}{x}\right)^7
			-\frac{\num{1398454401712471685447}}{\num{89357661481232687431680}}\left(1-\frac{1}{x}\right)^8+O((1-1/x)^9)\Bigr]\notag\\
			&-\frac{\Gamma(\frac{5}{12})\Gamma(\frac{11}{12})}{2\Gamma(\frac{4}{3})}\left(1-\frac{1}{x}\right)^{1/6}
			\Bigl[1-\frac{23}{192}\left(1-\frac{1}{x}\right)-\frac{\num{9071}}{\num{129024}}\left(1-\frac{1}{x}\right)^2
			-\frac{\num{1790683}}{\num{37158912}}\left(1-\frac{1}{x}\right)^3\notag\\
			&-\frac{\num{3349966483}}{\num{92748644352}}\left(1-\frac{1}{x}\right)^4
			-\frac{\num{2038392161807}}{\num{71230958862336}}\left(1-\frac{1}{x}\right)^5
			-\frac{\num{9176033135596199}}{\num{389775806894702592}}\left(1-\frac{1}{x}\right)^6\notag\\
			&-\frac{\num{744612558624442939}}{\num{37418477461891448832}}\left(1-\frac{1}{x}\right)^7
			-\frac{\num{246696882504200911723}}{\num{14368695345366316351488}}\left(1-\frac{1}{x}\right)^8+O((1-1/x)^9)\Bigr]\notag\\
			&-\frac{\sqrt{6}}{2}\Bigl[1-\frac{13}{140}\left(1-\frac{1}{x}\right)-\frac{\num{4391}}{\num{80080}}\left(1-\frac{1}{x}\right)^2
			-\frac{\num{3991667}}{\num{103463360}}\left(1-\frac{1}{x}\right)^3
			-\frac{\num{281688373}}{\num{9518629120}}\left(1-\frac{1}{x}\right)^4\notag\\
			&-\frac{\num{35576577679}}{\num{1488216970240}}\left(1-\frac{1}{x}\right)^5
			-\frac{\num{506348928456123}}{\num{25329452833484800}}\left(1-\frac{1}{x}\right)^6
			-\frac{\num{3060959470224584451}}{\num{178623301381734809600}}\left(1-\frac{1}{x}\right)^7\notag\\
			&-\frac{\num{3518168866000836642831}}{\num{235068264618363009433600}}\left(1-\frac{1}{x}\right)^8+O((1-1/x)^9)\Bigr].
		\end{align}

	\subsection{$x\to\infty$ (Head-on limit)}
		The exact expression (\ref{dphiinfinity}) can be expanded as
		\begin{align}
			\Delta\phi(x)=&\frac{3^{1/4}\pi^{3/2}}{2\Gamma^2(\frac{3}{4})}x^{-1/4}\Bigl[1+\frac{13}{72x}+\frac{\num{3265}}{\num{31104}x^2}
			+\frac{\num{508025}}{\num{6718464}x^3}+\frac{\num{809285165}}{\num{13544423424}x^4}
			+\frac{\num{436386280213}}{\num{8776786378752}x^5}\notag\\
			&+\frac{\num{254806596628207}}{\num{5958184124547072}x^6}+\frac{\num{16146758993098825}}{\num{428989256967389184}x^7}
			+\frac{\num{24975259828619413625}}{\num{741293436039648509952}x^8}+O(x^{-9})\Bigr]\notag\\
			&+\frac{\pi^{3/2}}{3^{1/4}\Gamma^2(\frac{1}{4})}x^{-3/4}\Bigl[1+\frac{85}{216x}+\frac{\num{38689}}{\num{155520}x^2}
			+\frac{\num{6164851}}{\num{33592320}x^3}+\frac{\num{2545831211}}{\num{17414258688}x^4}
			+\frac{\num{152886717137}}{\num{1253826625536}x^5}\notag\\
			&+\frac{\num{738395403493517}}{\num{7041490329010176}x^6}
			+\frac{\num{10783612337956183}}{\num{116997070082015232}x^7}
			+\frac{\num{1226414944517358286735}}{\num{14893259033160210972672}x^8}+O(x^{-9})\Bigr]\notag\\
			&+\sqrt{\frac{8}{3x}}\Bigl[1+\frac{43}{135x}+\frac{\num{15049}}{\num{76545}x^2}+\frac{\num{2847527}}{\num{19702683}x^3}
			+\frac{\num{11457019}}{\num{99379467}x^4}+\frac{\num{63908692867}}{\num{662761665423}x^5}\notag\\
			&+\frac{\num{25939601278931}}{\num{311799238051275}x^6}
			+\frac{\num{12399203454788981}}{\num{169019171580564225}x^7}
			+\frac{\num{1860211459709064161}}{\num{28293809322586451265}x^8}+O(x^{-9})\Bigr].
		\end{align}
}}}

\printbibliography
\end{document}